\documentclass[%
 reprint, 
 superscriptaddress,
 amsmath,amssymb,
 aps, physrev,
floatfix,
]{revtex4-2}

\usepackage{graphicx}
\usepackage{dcolumn}
\usepackage{bm}
\usepackage[hidelinks]{hyperref}

\begin{document}

\preprint{APS/123-QED}

\title{\textbf{SPIRE: Spin-Resolved Particle-in-Cell with Radiative Corrections for Extreme Laser--Plasma Interactions}
}%

\author{Xian-Zhang Wu}
\affiliation{Department of Nuclear Science and Technology, Xi'an Jiaotong University, Xi'an, China}

\author{Huai-Hang Song}
\affiliation{School of Physics, Zhengzhou University, Henan, China}

\author{Wei-Min Wang}
 \email{weiminwang1@sjtu.edu.cn}
\affiliation{School of Physics and Astronomy, Shanghai Jiao Tong University, Shanghai, China}

\author{Yan-Fei Li}
 \email{liyanfei@xjtu.edu.cn}
\affiliation{Department of Nuclear Science and Technology, Xi'an Jiaotong University, Xi'an, China}

\date{\today}

\begin{abstract}
We present SPIRE, a spin- and photon-polarization-resolved particle-in-cell (PIC)
code for self-consistent simulations of laser–plasma interactions in the 
strong-field quantum-electrodynamic (QED) regime. The classical PIC framework 
combines a Yee-grid finite-difference time-domain Maxwell solver, a relativistic 
Boris particle pusher, shape-function particle–grid interpolation, Esirkepov 
charge-conserving current deposition, Maxwell-consistent laser boundary 
injection, and a relativistic moving window, with implementations in one, 
two, and three spatial dimensions. A vectorial Laguerre–Gaussian laser 
injector based on an angular-spectrum formulation enables simulations of 
tightly focused, nonparaxial optical orbital-angular-momentum modes. The 
strong-field QED module employs a semiclassical approach. Nonlinear 
Compton scattering and nonlinear Breit–Wheeler pair production are 
treated as stochastic processes using spin- and polarization-resolved 
QED probabilities within the local constant-field approximation. 
The orbital motion of charged particles is governed by the classical 
Lorentz equation, while their spin evolution includes stochastic 
transitions associated with photon emission, no-emission self-energy 
corrections, and Thomas–Bargmann–Michel–Telegdi precession with a 
field-dependent anomalous magnetic moment. Photon polarization is 
described by the Stokes vector and evolves through nonlinear Compton 
scattering, nonlinear Breit–Wheeler pair production, vacuum birefringence, 
and vacuum dichroism. The implementation is verified through a set of one-, 
two-, and three-dimensional(1D, 2D, and 3D) benchmark problems. 
Classical PIC results are cross-checked against the open-source code 
SMILEI for electrostatic expansion, laser-driven radiation, beam-driven 
wakefields, laser wakefield acceleration with a moving window, and vectorial 
Laguerre–Gaussian laser injection. The spin- and polarization-resolved QED 
modules reproduce published results for polarized electron generation, vacuum 
birefringence and dichroism, and dense polarized positron production. 
These benchmarks provide numerical verification of the coupled classical PIC, 
strong-field QED, spin, and photon-polarization modules implemented in SPIRE.

\end{abstract}

\maketitle

\newpage

\section{Introduction}

In extreme laser--plasma interactions, sufficiently strong electromagnetic fields 
can give rise to significant quantum electrodynamic (QED) effects~\cite{Piazza2012,Gonoskov2022,FEDOTOV2023}.
Under such conditions, particle spin and photon polarization 
become essential degrees of freedom in physical processes, including radiation 
reaction. We have developed a laser--plasma simulation code that incorporates 
spin- and polarization-resolved strong-field QED processes, together with radiative 
and vacuum-polarization corrections, while self-consistently modeling the nonlinear 
evolution of electromagnetic fields and the collective behavior of particles in 
extreme laser-plasma interactions. The code is intended to facilitate 
investigations of spin and polarization dynamics and radiation-reaction effects 
in these regimes.

This paper presents the physical models and numerical methods implemented in the 
code. Benchmark comparisons with the open-source code SMILEI provide verification 
of the classical particle-in-cell(PIC) implementation, while comparisons with published results are 
used to verify the spin- and polarization-resolved QED modules. 
The PIC component 
was developed with reference to the implementations in 
KLAPS~\cite{Wang2015,Wang2025} and SMILEI~\cite{DEROUILLAT2018351}, 
as well as selected algorithms from YUNIC~\cite{YUNIC} and SLIPs~\cite{Wan2023}. 
The QED algorithms were implemented with reference to those employed in 
CAIN~\cite{CAIN1} and YUNIC~\cite{YUNIC}.

\section{Laser--Plasma Interaction Simulations}


The inherent complexity of laser--plasma interactions poses a formidable challenge 
for analytical descriptions. Consequently, numerical simulations have become a 
robust and indispensable approach for investigating the diverse physical phenomena 
involved. In this study, we employ the PIC
method~\cite{Birdsall1991Plasma,RuhlPSCIntro}, in which electromagnetic fields are 
discretized on a coordinate grid and evolved using the finite-difference 
time-domain (FDTD) scheme to solve Maxwell's equations. Meanwhile, the plasma 
is represented by charged macroparticles, each treated as a ``particle cloud'' 
with a prescribed shape function and representing a specific number of physical 
particles.

The application of PIC methods to simulate laser-plasma interactions has a long 
history and is now widely utilized in prominent research areas such as electron 
beam acceleration, terahertz (THz) radiation, inertial confinement fusion (ICF), 
and the dynamics of polarized particle beams, underscoring its significant scientific 
and technical impact. Following the architecture of SMILEI, we have implemented a 
PIC code for laser–plasma interactions. Its classical module is cross-checked 
against SMILEI in Sec.~\ref{sec:benchmark cases}.

\subsection{Governing Equations}

The PIC method describes the self-consistent dynamics of
charged particles and electromagnetic fields on 1D, 2D, or 3D
spatial grids. In Gaussian units, the
governing equations consist of Maxwell's equations, the relativistic Lorentz
equation of motion, and the charge-continuity equation:
\begin{equation}
\begin{aligned}
\nabla\times\mathbf{E}
  &= -\frac{1}{c}\frac{\partial\mathbf{B}}{\partial t}, \\
\nabla\times\mathbf{B}
  &= \frac{1}{c}\frac{\partial\mathbf{E}}{\partial t}
     +\frac{4\pi}{c}\mathbf{J}, \\
\nabla\cdot\mathbf{E}
  &= 4\pi\rho, \\
\nabla\cdot\mathbf{B}
  &= 0, \\
\frac{\mathrm{d}\mathbf{p}}{\mathrm{d}t}
  &= q\left(
       \mathbf{E}
       +\frac{\mathbf{v}}{c}\times\mathbf{B}
     \right), \\
\frac{\partial\rho}{\partial t}
  &+\nabla\cdot\mathbf{J}=0.
\end{aligned}
\label{eq:governing_equations}
\end{equation}
\subsection{Units}

To enhance computational efficiency and ensure numerical stability, 
physical quantities in the PIC scheme are normalized using the 
elementary charge $q_e$, the electron mass $m_e$, the speed of 
light $c$, and a reference frequency $\omega_r$ as fundamental 
base units. The selection of $\omega_r$ depends on the primary 
focus of the simulation: for laser-driven processes, $\omega_r$ 
is typically defined as the laser angular frequency ($\omega_0$), 
while for plasma-dominated interactions, it represents the plasma 
frequency ($\omega_p$). Following these basic units and physical
relations, the normalization of time ($t_r$), electric field 
intensity ($E_r$), magnetic induction ($B_r$), current density 
($J_r$), critical plasma density ($n_r$), wavelength ($\lambda_r$), 
velocity ($V_r$), and momentum ($P_r$) is as follows:

\begin{equation}
\begin{aligned}
t_r &= \frac{1}{\omega_r},
& L_r &= \frac{c}{\omega_r},
& v_r &= c, \\
E_r &= \frac{m_e c \omega_r}{e},
& B_r &= \frac{m_e c \omega_r}{e},
& p_r &= m_e c, \\
n_r &= \frac{m_e \omega_r^2}{4\pi e^2},
& J_r &= c e n_r,
& q_r &= e,\quad m_r=m_e .
\end{aligned}
\end{equation}

\subsection{Solving Electromagnetic Fields}


In our PIC framework, the Yee lattice~\cite{1966Numerical} is 
employed to solve for the electromagnetic fields. 
Within the Yee grid, electric and magnetic field 
components are distributed in a spatially staggered 
manner to naturally accommodate the curl operators 
in Maxwell's equations, thereby ensuring high 
numerical precision and stability. For temporal 
field advancement, we use the 
FDTD method~\cite{2005Computational}, 
which solves Maxwell's equations in 
the time domain by approximating partial 
derivatives with finite differences.

\begin{figure}
    \centering
    \includegraphics[width=\columnwidth]{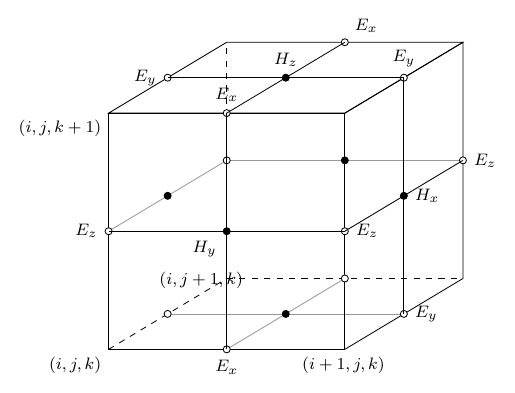}
    \caption{Schematic representation of the spatial distribution of electric and magnetic field components within the Yee grid. White open circles and black solid circles denote the electric and magnetic field components, respectively.}
    \label{figure:Yee}
\end{figure}

As illustrated in Fig.~\ref{figure:Yee}, the Yee lattice allocates the electric-field components to the centers of grid edges oriented along their respective coordinate directions, while the magnetic-field components are located at the centers of the grid faces normal to their respective directions. This arrangement is designed for the discrete evaluation of the curl operators in Maxwell's equations. Because Faraday's and Ampère's laws couple the curls of the electric and magnetic fields, the two sets of field components are spatially staggered so that they effectively ``encircle'' one another.

By offsetting these components by half a grid spacing, the finite-difference scheme captures the geometric structure of the curl. For example, the update of $B_x$ directly uses the surrounding $E_y$ and $E_z$ components. Combined with a half-step temporal staggering, in which the electric and magnetic fields are evaluated at integer and half-integer time levels, respectively, this arrangement yields second-order accuracy. Equation~\ref{equation:eq2-3} is discretized on the Yee grid as Eq.~\ref{equation:eq2-4}:

\begin{equation}
    \frac{\partial{B_x}}{\partial t}= \frac{\partial{E_y}}{\partial z}-\frac{\partial{E_z}}{\partial y}
    \label{equation:eq2-3}
\end{equation}

\begin{equation}
\begin{aligned}
&\frac{B_x^{n+1/2}\!\left(i,j+\frac{1}{2},k+\frac{1}{2}\right)
-B_x^{n-1/2}\!\left(i,j+\frac{1}{2},k+\frac{1}{2}\right)}{\Delta t} \\
={}&\frac{E_y^n\!\left(i,j+\frac{1}{2},k+1\right)
-E_y^n\!\left(i,j+\frac{1}{2},k\right)}{\Delta z} \\
&-\frac{E_z^n\!\left(i,j+1,k+\frac{1}{2}\right)
-E_z^n\!\left(i,j,k+\frac{1}{2}\right)}{\Delta y}.
\end{aligned}
\label{equation:eq2-4}
\end{equation}

To ensure that the electromagnetic fields are accurately resolved and that their variations remain sufficiently gradual within a single spatial step, the grid spacing must be a small fraction of the wavelength. Furthermore, to maintain numerical stability, the spatial and temporal steps must satisfy the Courant-Friedrichs-Lewy (CFL) condition \cite{Courant1928}:

\begin{equation}
    c \Delta t \leq \left[ \sum_\mu \Delta \mu^{-2} \right]^{-1/2},
    \label{equation:eq2-5}
\end{equation}
where $\mu = (x, y, z)$ denotes the different spatial directions resolved in the simulation. 
For a 3D uniform grid with $\Delta x = \Delta y = \Delta z = \Delta$, this condition simplifies to $\Delta t \leq \Delta / (c \sqrt{3})$.

\subsection{Particle Pusher}

The motion of a charged particle in electromagnetic fields is governed by the
relativistic Lorentz equation. In Gaussian units, the momentum and position
equations are
\begin{equation}
\frac{\mathrm{d}\mathbf{p}}{\mathrm{d}t}
=
q\left(
\mathbf{E}
+\frac{\mathbf{v}}{c}\times\mathbf{B}
\right),
\qquad
\frac{\mathrm{d}\mathbf{r}}{\mathrm{d}t}
=
\mathbf{v},
\label{eq:lorentz_equations}
\end{equation}
where
\begin{equation}
\mathbf{p}=\gamma m\mathbf{v},
\qquad
\gamma=
\frac{1}{\sqrt{1-v^2/c^2}}.
\label{eq:momentum_velocity_relation}
\end{equation}

For numerical integration, we introduce the momentum per unit mass,
\begin{equation}
\mathbf{u}
\equiv
\frac{\mathbf{p}}{m}
=
\gamma\mathbf{v},
\qquad
\gamma
=
\sqrt{1+\frac{u^2}{c^2}}.
\label{eq:proper_velocity}
\end{equation}
The discretized equations of motion can then be written as
\begin{equation}
\frac{\mathbf{u}^{n+1}-\mathbf{u}^{n}}{\Delta t}
=
\frac{q}{m}
\left(
\mathbf{E}_p^{\,n+\frac{1}{2}}
+
\frac{\mathbf{u}^{\,n+\frac{1}{2}}}
     {\gamma^{\,n+\frac{1}{2}}c}
\times
\mathbf{B}_p^{\,n+\frac{1}{2}}
\right),
\label{eq:discrete_momentum}
\end{equation}
and
\begin{equation}
\frac{
\mathbf{r}_p^{\,n+\frac{3}{2}}
-
\mathbf{r}_p^{\,n+\frac{1}{2}}
}{\Delta t}
=
\frac{\mathbf{u}^{n+1}}{\gamma^{n+1}}.
\label{eq:discrete_position}
\end{equation}

Here, $\mathbf{r}_p$ denotes the particle position, $\mathbf{u}$ is the
momentum per unit mass, and $\gamma$ is the Lorentz factor. The quantities
$\mathbf{E}_p$ and $\mathbf{B}_p$ are the electric and magnetic fields
interpolated to the particle position. The charge-to-mass ratio is denoted by
$q/m$.

We integrate these equations using the relativistic Boris algorithm
\cite{Boris1970PlasmaConf}. The method separates the electric-field
acceleration from the magnetic-field rotation. First, the electric field
advances the momentum by half a time step:
\begin{equation}
\mathbf{u}^{-}
=
\mathbf{u}^{n}
+
\frac{q\Delta t}{2m}
\mathbf{E}_p^{\,n+\frac{1}{2}}.
\label{eq:boris_first_electric_push}
\end{equation}
The corresponding Lorentz factor is
\begin{equation}
\gamma^{-}
=
\sqrt{1+\frac{|\mathbf{u}^{-}|^2}{c^2}}.
\label{eq:boris_gamma_minus}
\end{equation}
The magnetic rotation is then performed according to
\begin{equation}
\begin{aligned}
\mathbf{t}
&=
\frac{q\Delta t}
     {2m\gamma^{-}c}
\mathbf{B}_p^{\,n+\frac{1}{2}},
\\
\mathbf{s}
&=
\frac{2\mathbf{t}}{1+t^2},
\\
\mathbf{u}^{\prime}
&=
\mathbf{u}^{-}
+
\mathbf{u}^{-}\times\mathbf{t},
\\
\mathbf{u}^{+}
&=
\mathbf{u}^{-}
+
\mathbf{u}^{\prime}\times\mathbf{s}.
\end{aligned}
\label{eq:boris_rotation}
\end{equation}
Finally, the second half of the electric-field acceleration gives
\begin{equation}
\mathbf{u}^{n+1}
=
\mathbf{u}^{+}
+
\frac{q\Delta t}{2m}
\mathbf{E}_p^{\,n+\frac{1}{2}}.
\label{eq:boris_second_electric_push}
\end{equation}

\begin{figure}[t]
  \centering
  \includegraphics[width=\columnwidth]
    {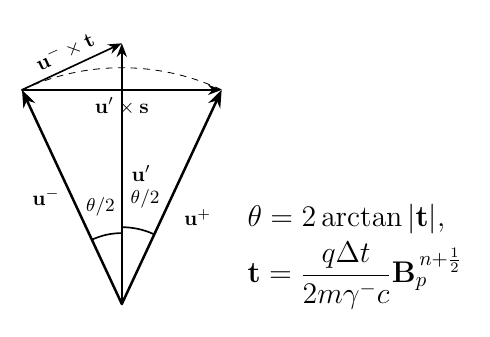}
  \caption{
    Geometric representation of the magnetic-rotation step in the Boris
    algorithm. The vector $\mathbf{u}^{-}$ is the momentum per unit mass
    after the first half-step electric-field acceleration,
    $\mathbf{u}^{\prime}$ is an intermediate vector, and
    $\mathbf{u}^{+}$ is the vector after the magnetic-field rotation.
    The angle $\theta$ denotes the numerical rotation angle.
  }
  \label{fig:boris_rotation}
\end{figure}

Figure~\ref{fig:boris_rotation} illustrates the magnetic-rotation step in the
plane perpendicular to $\mathbf{B}$. Because a magnetic field performs no work,
the magnitude of the momentum is conserved during this step:
$|\mathbf{u}^{+}|=|\mathbf{u}^{-}|$. The numerical rotation angle is
\begin{equation}
\theta
=
2\arctan|\mathbf{t}|
=
2\arctan\left(
\frac{|q|B\Delta t}{2m\gamma^{-}c}
\right).
\label{eq:boris_rotation_angle}
\end{equation}
\subsection{Field Interpolation and Current Deposition}

To enhance numerical precision and stability, macroparticles are assigned spatial shape functions. Thus, each simulated particle is treated not as a point on the grid but as a ``particle cloud'' extending over several cells~\cite{Birdsall1991Plasma}. The fields acting on a macroparticle are obtained by interpolating the grid-based field components with this shape function. Conversely, the particle charge and current are deposited onto the grid using the same function. Using a consistent shape function for interpolation and deposition prevents unphysical self-forces.

The shape function of a macroparticle, $S(\mathbf{x})$, is constructed as the product of 
1D shape functions $s(x)$ along each axis. In 3D space, it is

\begin{equation}
S(x,y,z)={s}^{(n)}(x) \cdot {s}^{(n)}(y) \cdot {s}^{(n)}(z),
\end{equation}
here, $n$ denotes the order of the shape function, representing the degree of interpolation or deposition. A zero-order shape function, $s^{(0)}(x) = \delta(x)$, where $\delta(x)$ is the Dirac delta function, characterizes the particle as a discrete point. Higher-order shape functions can be derived through the following recurrence relation \cite{ABE1986247}:

\begin{equation}
\begin{aligned}
s^{(n)}(x) & =\Delta x^{-1} P(x) \otimes s^{(n-1)}(x) \\
& \equiv \Delta x^{-1} \int_{-\infty}^{\infty} d x^{\prime} P\left(x^{\prime}-x\right) s^{(n-1)}\left(x^{\prime}\right),
\end{aligned}
\end{equation}

where $P(x)=1$ for $|x| \leq \frac{1}{2} \Delta x$ and $P(x)=0$ otherwise. Based on the recurrence relation above, the explicitly normalized shape functions $\hat{s}^{(n)} = \Delta x s^{(n)}$ can be written compactly by introducing $\zeta=|x|/\Delta x$:

\begin{equation}
\begin{aligned}
& \hat{s}^{(0)}(x)=\Delta x \delta(x), \\
& \hat{s}^{(1)}(x)=
\begin{cases}1, & \zeta\leq\frac12,\\ 0, & \text{otherwise},\end{cases} \\
& \hat{s}^{(2)}(x)=
\begin{cases}1-\zeta, & \zeta\leq1,\\ 0, & \text{otherwise},\end{cases} \\
& \hat{s}^{(3)}(x)=
\begin{cases}
\frac34-\zeta^2, & \zeta\leq\frac12,\\
\frac98\left(1-\frac23\zeta\right)^2,
  & \frac12<\zeta\leq\frac32,\\
0, & \text{otherwise},
\end{cases} \\
& \hat{s}^{(4)}(x)=
\begin{cases}
\frac23\left(1-\frac32\zeta^2+\frac34\zeta^3\right),
  & \zeta\leq1,\\
\frac43\left(1-\frac12\zeta\right)^3,
  & 1<\zeta\leq2,\\
0, & \text{otherwise}.
\end{cases}
\end{aligned}
\end{equation}

\begin{figure}
    \centering
    \includegraphics[width=\columnwidth]{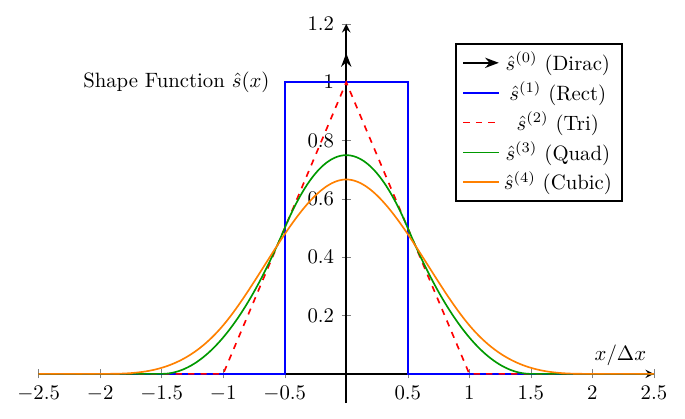}
    \caption{Macroparticle shape functions of different orders. The horizontal axis denotes the distance from the particle center, and the vertical axis denotes the shape-function value.}
    \label{figure:shape_function}
\end{figure}

For the 1D case, the effective electric or magnetic field at the particle position is calculated as
\begin{equation}
F_p = \sum_{i} F_i \hat s^{(n+1)} (x_p - x_i)
\end{equation} 
where $i$ denotes the grid index, $F$ represents the electric or magnetic field components, and $x_p$ and $x_i$ are the particle and grid positions, respectively.

Similarly, the macroparticle charge and current densities are deposited onto the grid using the same shape function:
\begin{equation}
\rho_i = \sum_{p} q_p \hat s^{(n+1)} (x_p - x_i)
\end{equation}

\subsection{Current Deposition Algorithm}

To maintain charge conservation, the current density generated by particle motion must be evaluated accurately because it is a source term in the electric-field update. Since macroparticles have finite spatial extent, the current-deposition scheme must account for both the particle shape functions and the grid geometry. We employ the charge-conserving current-deposition method of Esirkepov~\cite{ESIRKEPOV2001144}:

\begin{equation}
\begin{aligned}
\mathcal{J}_{i+1, j, k}^x-\mathcal{J}_{i, j, k}^x & =-\mathrm{Q} \frac{d x}{d t} \mathrm{~W}^{x}(i, j, k), \\
\mathcal{J}_{i, j+1, k}^y-\mathcal{J}_{i, j, k}^y & =-\mathrm{Q} \frac{d y}{d t} \mathrm{~W}^{y}(i, j, k), \\
\mathcal{J}_{i, j, k+1}^z-\mathcal{J}_{i, j, k}^z & =-\mathrm{Q} \frac{d z}{d t} \mathrm{~W}^{z}(i, j, k),
\end{aligned}
\end{equation}

Here, $\mathrm{Q}$ denotes the macroparticle charge, and $\mathrm{W}$ is a weighting factor that depends on the particle positions at the previous and subsequent time levels. The indices $i$, $j$, and $k$ denote 3D grid indices. Using cyclic coordinate triples $(a,b,c)\in\{(x,y,z),(y,x,z),(z,x,y)\}$ and the corresponding grid indices $(i_a,i_b,i_c)$, the three weighting factors can be written as

\begin{equation}
\begin{aligned}
W^a(i_a,i_b,i_c)={}&\mathrm{DS}^a(i_a)
\Bigl[\mathrm{S0}^b(i_b)\mathrm{S0}^c(i_c)\\
&+\frac12\mathrm{DS}^b(i_b)\mathrm{S0}^c(i_c)
+\frac12\mathrm{S0}^b(i_b)\mathrm{DS}^c(i_c)\\
&+\frac13\mathrm{DS}^b(i_b)\mathrm{DS}^c(i_c)\Bigr].
\end{aligned}
\end{equation}

Here, $\mathrm{S0}$ and $\mathrm{S1}$ denote the values of the shape function at the particle's positions at the previous and subsequent time steps, respectively, following the methodology described in the field interpolation section. The term $\mathrm{DS} = \mathrm{S1} - \mathrm{S0}$ represents the increment of the shape function.

\subsection{Laser Boundary Injection}

In PIC simulations, laser pulses are typically introduced into the computational domain through the boundary injection method. This approach involves prescribing appropriate electromagnetic field boundary conditions at the domain boundaries to simulate the propagation of laser pulses into the interior. The detailed implementation of laser injection and absorption at the boundaries, derived from Maxwell's equations, is as follows \cite{RuhlPSCIntro}.

Maxwell's equations can be reformulated as:

\begin{equation}
\begin{aligned}
& \partial_t\left(E_y \pm B_z\right) \pm \partial_x\left(E_y \pm B_z\right)= \pm \partial_y E_x+\partial_z B_x-j_y \\
& \partial_t\left(E_z \mp B_y\right) \pm \partial_x\left(E_z \mp B_y\right)= \pm \partial_z E_x-\partial_y B_x-j_z
\end{aligned}
\label{equation:eq2-16}
\end{equation}

The reformulated equations above can be used to define absorbing (radiating) boundary conditions. If the $E_x$ and $B_x$ components vanish, Eq.~\ref{equation:eq2-16} describes traveling waves propagating along the $x$ axis in vacuum. These relations yield the following expressions for $S$-polarized waves $(E_y,B_z)$ and $P$-polarized waves $(E_z,B_y)$ propagating in the positive or negative $x$ direction:

\begin{equation}
\begin{aligned}
2S^n_{j,k+\frac12,l}
={}&(E_y^n)_{j,k+\frac12,l}
\pm(B_z^n)_{j,k+\frac12,l} \\
={}&(E_y^{n-\frac12})_{j,k+\frac12,l} \\
&+\frac{\Delta t}{2\Delta z}
\Bigl[(B_x^n)_{j,k+\frac12,l+\frac12}
      -(B_x^n)_{j,k+\frac12,l-\frac12}\Bigr] \\
&-\frac{\Delta t}{2\Delta x}
\Bigl[(B_z^n)_{j+\frac12,k+\frac12,l}
      -(B_z^n)_{j-\frac12,k+\frac12,l}\Bigr] \\
&\pm\frac12
\Bigl[(B_z^n)_{j+\frac12,k+\frac12,l}
      +(B_z^n)_{j-\frac12,k+\frac12,l}\Bigr] \\
&-\frac{\Delta t}{2}(j_y^n)_{j,k+\frac12,l}.
\end{aligned}
\end{equation}

\begin{equation}
\begin{aligned}
2P^n_{j,k,l+\frac12}
={}&(E_z^n)_{j,k,l+\frac12}
\mp(B_y^n)_{j,k,l+\frac12} \\
={}&(E_z^{n-\frac12})_{j,k,l+\frac12} \\
&+\frac{\Delta t}{2\Delta x}
\Bigl[(B_y^n)_{j+\frac12,k,l+\frac12}
      -(B_y^n)_{j-\frac12,k,l+\frac12}\Bigr] \\
&-\frac{\Delta t}{2\Delta y}
\Bigl[(B_x^n)_{j,k+\frac12,l+\frac12}
      -(B_x^n)_{j,k-\frac12,l+\frac12}\Bigr] \\
&\mp\frac12
\Bigl[(B_y^n)_{j+\frac12,k,l+\frac12}
      +(B_y^n)_{j-\frac12,k,l+\frac12}\Bigr] \\
&-\frac{\Delta t}{2}(j_z^n)_{j,k,l+\frac12}.
\end{aligned}
\end{equation}

By reformulating these expressions, the target laser amplitude can be assigned to the electromagnetic field components, enabling the spatiotemporal evolution of the laser pulse through the numerical solution of Maxwell's equations. For forward-propagating laser injection at the left boundary, the equations are rewritten as:

\begin{equation}
\begin{aligned}
\left(B_z\right)_{j-\frac{1}{2}, k+\frac{1}{2}, l}^n= & \frac{1}{1+\frac{\Delta t}{\Delta x}}\left[4 S_{j, k+\frac{1}{2}, l}^n-2\left(E_y\right)_{j, k+\frac{1}{2}, l}^{n-\frac{1}{2}}\right. \\
& -\frac{\Delta t}{\Delta z}\left(\left(B_x\right)_{j, k+\frac{1}{2}, l+\frac{1}{2}}^n-\left(B_x\right)_{j, k+\frac{1}{2}, l-\frac{1}{2}}^n\right) \\
& -\left(1-\frac{\Delta t}{\Delta x}\right)\left(B_z\right)_{j+\frac{1}{2}, k+\frac{1}{2}, l}^n \\
& \left.-\Delta t\left(j_y\right)_{j, k+\frac{1}{2}, l}^n\right]
\end{aligned}
\end{equation}

\begin{align}
\left(B_y\right)_{j-\frac{1}{2}, k, l+\frac{1}{2}}^n= & \frac{1}{1+\frac{\Delta t}{\Delta x}}\left[-4 P_{j, l, k+\frac{1}{2}}^n+2\left(E_z\right)_{j, k, l+\frac{1}{2}}^{n-\frac{1}{2}}\right. \nonumber  \\
+ & \frac{\Delta t}{\Delta y}\left(\left(B_x\right)_{j, k+\frac{1}{2}, l+\frac{1}{2}}^n-\left(B_x\right)_{j, k-\frac{1}{2}, l+\frac{1}{2}}^n\right) \nonumber \\
- & \left(1-\frac{\Delta t}{\Delta x}\right)\left(B_y\right)_{j+\frac{1}{2}, k, l+\frac{1}{2}}^n \nonumber \\
+ & \left.\Delta t\left(j_z\right)_{j, k, l+\frac{1}{2}}^n\right] .
\end{align}

In addition to specifying the wavelength, pulse duration, and intensity, the laser mode—which characterizes the spatial distribution of the electromagnetic field—must also be defined. The following expression represents a short-pulse Gaussian beam in its fundamental mode, taking an S-polarized wave as an example \cite{Shen2023StrongField}:

\begin{equation}
\begin{aligned}
E(y,z,x,t)={}&E_0
 \exp\!\left[-\frac{y^2+z^2}{w^2(x)}\right] \\
&\times\exp\!\left[-\frac{(t-x/c)^2}{\tau^2}\right]\\
&\times\cos[\omega_0(t-x/c)+\phi_0].
\end{aligned}
\end{equation}

Here, $\exp \left[-(y^2+z^2)/w^2(x)\right]$ characterizes the transverse Gaussian profile of the laser beam perpendicular to the $x$-axis. The beam radius is defined as $w(x)=w_0 \sqrt{1+\left(x/z_R\right)^2}$, where $z_R=\pi w_0^2/\lambda$ is the Rayleigh length, $w_0$ is the beam waist radius at the focal plane, and $\lambda$ is the laser wavelength. The term $\exp \left[-(t-x/c)^2/\tau^2\right]$ represents the temporal Gaussian envelope of the short pulse, in which $t-x/c$ denotes the retarded time for propagation along the $x$-axis and $\tau$ denotes the $1/e$ pulse duration. The term $\cos \left[\omega_0(t-x/c)+\phi_0\right]$ describes the carrier oscillation of the laser field, with $\omega_0$ being the carrier frequency and $\phi_0$ the initial phase.

\subsection{Vectorial Laguerre--Gaussian Laser Injection}
\label{sec:lg_injection}

A tightly focused Laguerre--Gaussian (LG) laser pulse is injected into
the simulation domain using a vectorial angular-spectrum formulation
combined with a total-field/scattered-field (TF/SF) boundary
\cite{Thiele2016TightFocus,Capoglu2008FocusedFDTD,
Charbonnet2025FocusedPIC}. The
incident mode is specified in the pupil plane by its radial index $p$,
azimuthal index $\ell$, numerical aperture $\mathrm{NA}$, pupil filling
factor $f_{\mathrm{p}}$, and polarization state. In transverse
wave-number space, the scalar pupil amplitude is defined as
\begin{equation}
\begin{aligned}
\widetilde{u}_{p\ell}(k_\perp,\varphi_k)
={}&\left(\sqrt{2}\rho\right)^{|\ell|}
L_p^{|\ell|}\!\left(2\rho^2\right)\\
&\quad\times\exp\!\left(-\rho^2+i\ell\varphi_k\right)
\sqrt{\cos\theta},
\label{eq:lg_pupil}
\end{aligned}
\end{equation}
where
\begin{equation}
\rho
=
\frac{\sin\theta}{\mathrm{NA}\,f_{\mathrm{p}}},
\qquad
\sin\theta
=
\frac{k_\perp}{k_0},
\qquad
k_\perp
=
\sqrt{k_y^2+k_z^2}.
\label{eq:lg_coordinates}
\end{equation}
Here, $L_p^{|\ell|}$ denotes the generalized Laguerre polynomial. The
factor $\exp(i\ell\varphi_k)$ introduces an azimuthal phase winding of
$2\pi\ell$, corresponding to an orbital angular momentum of
$\ell\hbar$ per photon in the paraxial limit
\cite{Allen1992OAM}. Spectral components outside the prescribed
numerical aperture,
\begin{equation}
k_\perp > k_0\mathrm{NA},
\end{equation}
are discarded.

For each propagating angular-spectrum component, the prescribed input
polarization is rotated into the local propagation frame. The
longitudinal electric-field component is determined by imposing the
transversality condition
\begin{equation}
\mathbf{k}\cdot\widetilde{\mathbf{E}}=0.
\label{eq:continuous_transversality}
\end{equation}
The corresponding magnetic field is obtained from the spectral Maxwell
relation
\begin{equation}
\widetilde{\mathbf{B}}
=
\frac{\mathbf{k}\times\widetilde{\mathbf{E}}}{\omega}.
\label{eq:continuous_magnetic_field}
\end{equation}
Consequently, the generated field includes the nonparaxial longitudinal
components and polarization mixing required for a tightly focused beam,
consistently with vectorial diffraction theory
\cite{RichardsWolf1959,KimLee1999NonparaxialLG,
Gonzalez2018MaxwellConsistent}. Zero padding is applied in the
transverse Fourier plane to improve the angular-spectrum resolution and
reduce wrap-around errors. An inverse 2D fast Fourier
transform is subsequently used to construct all six complex
electromagnetic-field components on the injection plane. These
components are sampled directly at their corresponding staggered
Yee-grid positions.

When the discrete-dispersion option is enabled, the continuous wave
vector and frequency are replaced by their modified Yee-grid
counterparts,
\begin{equation}
\widetilde{k}_i
=
\frac{2}{\Delta i}
\sin\left(\frac{k_i\Delta i}{2}\right),
\qquad
i\in\{x,y,z\},
\label{eq:modified_wavevector}
\end{equation}
and
\begin{equation}
\widetilde{\omega}
=
\frac{2}{\Delta t}
\sin\left(\frac{\omega\Delta t}{2}\right).
\label{eq:modified_frequency}
\end{equation}
The longitudinal wave number $k_x$ is then determined from the discrete
vacuum dispersion relation
\begin{equation}
\widetilde{\omega}^{\,2}
=
c^2
\left(
\widetilde{k}_x^2+
\widetilde{k}_y^2+
\widetilde{k}_z^2
\right).
\label{eq:yee_dispersion}
\end{equation}
The electric field is projected onto the discrete transverse subspace,
such that
\begin{equation}
\widetilde{\mathbf{k}}
\cdot
\widetilde{\mathbf{E}}
=0,
\label{eq:discrete_transversality}
\end{equation}
and the magnetic field is constructed from the corresponding discrete
electric--magnetic eigenmode relation. This procedure reduces
spurious backward-propagating fields caused by a mismatch between a
continuously constructed laser mode and the numerical dispersion of
the Yee Maxwell solver \cite{Yee1966}.

The resulting incident field is coupled to the PIC Maxwell solver
through a single-plane TF/SF formulation
\cite{UmashankarTaflove1982}. Complex field templates are generated on
both sides of the TF/SF interface and distributed among the MPI
subdomains intersecting the injection plane. During the simulation,
the instantaneous incident fields are reconstructed according to
\begin{equation}
\mathbf{F}_{\mathrm{inc}}(\mathbf{x},t)
=
g(t)\,
\operatorname{Re}
\left[
\widetilde{\mathbf{F}}(\mathbf{x})
\exp\left(-i\omega_0t+i\phi_0\right)
\right],
\label{eq:incident_field_reconstruction}
\end{equation}
where $\mathbf{F}$ represents either the electric or magnetic field,
$\omega_0$ is the carrier frequency, $\phi_0$ is the initial phase,
and $g(t)$ is the temporal envelope.
Corrections derived from the incident tangential magnetic field are
applied during the electric-field update, whereas corrections derived
from the incident tangential electric field are applied during the
magnetic-field update. Since only the precomputed complex
injection-plane templates are retained, the memory requirement is
independent of the laser-pulse duration.

The implementation additionally provides a frequency-resolved
broadband mode. In this mode, the transverse angular spectrum at the
carrier-frequency injection plane is retained, while the longitudinal
wave number, discrete polarization projection, and electric--magnetic
relation are recomputed independently for each temporal-frequency bin.
This treatment improves the consistency of short-pulse injection with
the frequency-dependent numerical dispersion of the Maxwell solver.
This frequency-resolved construction is particularly important for
ultrashort and tightly focused pulses, for which spatial and temporal
spectral components cannot generally be treated independently
\cite{Capoglu2008FocusedFDTD,Charbonnet2025FocusedPIC}.

\subsection{Benchmark Cases}
\label{sec:benchmark cases}
\subsubsection{1D Case: Electrostatic Repulsion}

\begin{figure}
    \centering
    \includegraphics[width=\columnwidth]{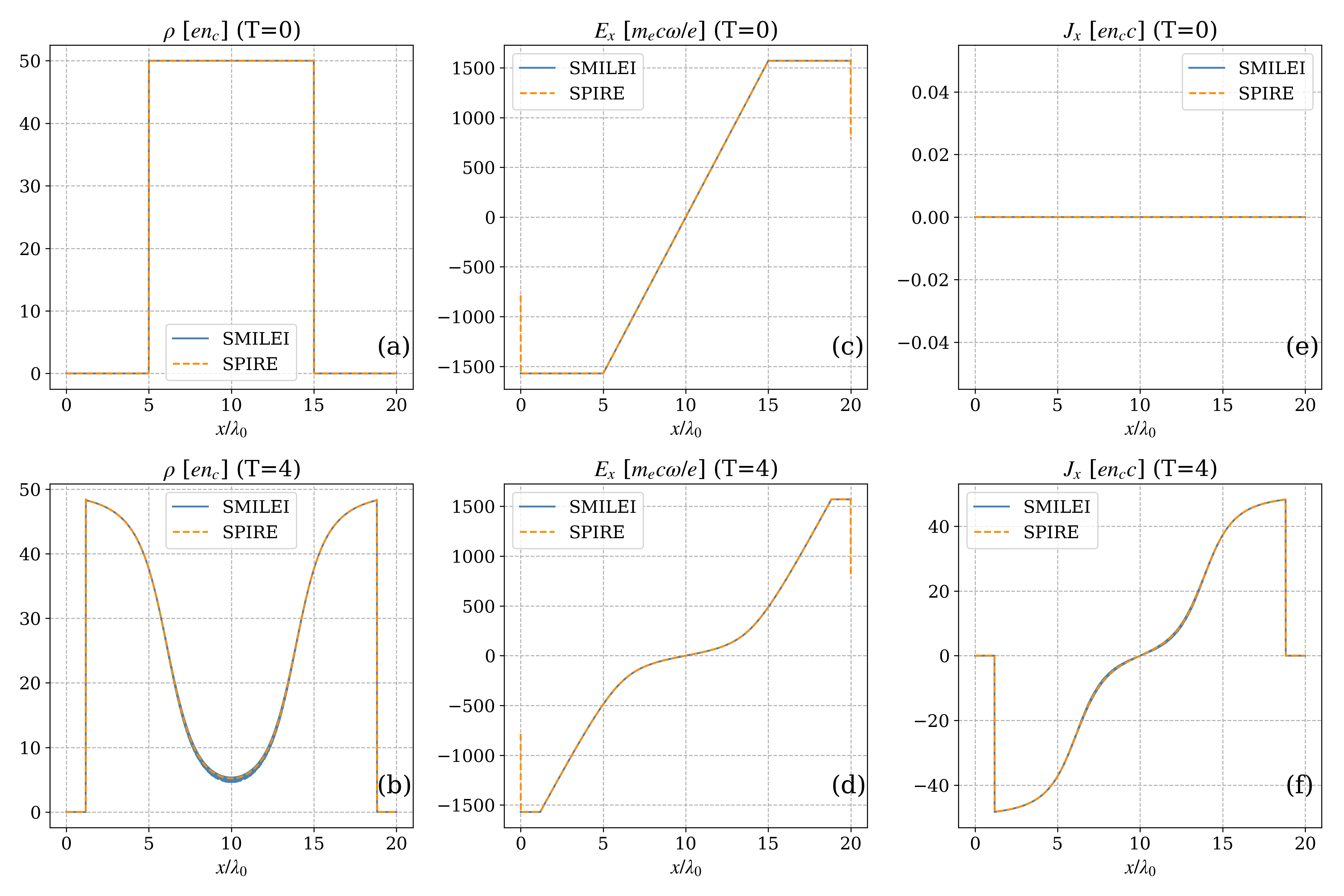}
    \caption{Comparison of results for the 1D electrostatic repulsion benchmark: (a)–(b) charge density distribution; (c)–(d) longitudinal electric field $E_x$; (e)–(f) current density $J_x$. The orange dashed lines represent the results from SPIRE, while the blue solid lines denote the results from SMILEI.}
    \label{fig:1d_benchmark}
\end{figure}

In this benchmark, an ensemble of initially stationary protons with a temperature of 0~eV expands in one dimension under its electrostatic self-field. A uniform proton density of $n_p=50n_c$, where $n_c=m_e\omega^2/(4\pi e^2)$ is the critical density in Gaussian units, is initialized in the region $5\lambda_L<x<15\lambda_L$. The domain length is $20\lambda_L$ and is resolved by 5120 cells with 100 macroparticles per cell. Absorbing boundary conditions are applied to both particles and fields, and second-order interpolation is used. Because the initial charge density is nonzero, Poisson's equation is solved before time integration to initialize the electric field. Figure~\ref{fig:1d_benchmark} compares the charge density, longitudinal electric field $E_x$, and current density $J_x$.

\subsubsection{2D Benchmark: THz Radiation from a Thin Plasma Layer}

An ultrashort intense laser pulse obliquely incident on a thin, underdense plasma layer—with a thickness on the order of the plasma wavelength ($L \sim \lambda_p$)—drives a transient net current via the ponderomotive force. In this thin-target regime, the displacement current fails to fully cancel the plasma current. Furthermore, at the plasma frequency $\omega_p$, the plasma skin depth $k_p^{-1}$ is comparable to the layer thickness $L$, allowing the generated radiation to penetrate the plasma and propagate into the vacuum\cite{PhysRevE.77.046405}.

In the corresponding 2D PIC simulation, the parameters are meticulously chosen to replicate this physical scenario. The large-scale simulation domain spans $300\lambda \times 300\lambda$, with a spatial resolution of 16 cells per wavelength and a temporal resolution of 32 steps per laser period. The calculation is performed using a $16 \times 16$ domain decomposition over a total duration of 300 cycles. The plasma thin layer is modeled as a rectangular slab centered at $(150\lambda, 150\lambda)$ with a length of $200\lambda$, a thickness of $25\lambda$, and a tilt angle of 50$^\circ$. The plasma density is set to $n = 0.0025n_c$.

The laser pulse is S-polarized with a peak normalized amplitude $a_0 \equiv |e|E_0/(m_e c \omega_0)= 0.5$ and a duration of 10 cycles. It is focused at $(x, y) = (150\lambda, 150\lambda)$ with a beam waist $w_L = 60\lambda$. The temporal envelope follows a $\sin^2$ profile: $a_L = a_0 \sin^2 [\pi(x-ct)/(10 \lambda_L)]$, where the pulse is injected from the left boundary starting at the fifth laser cycle.

\begin{figure}[htbp]
    \centering
    \includegraphics[width=\columnwidth]{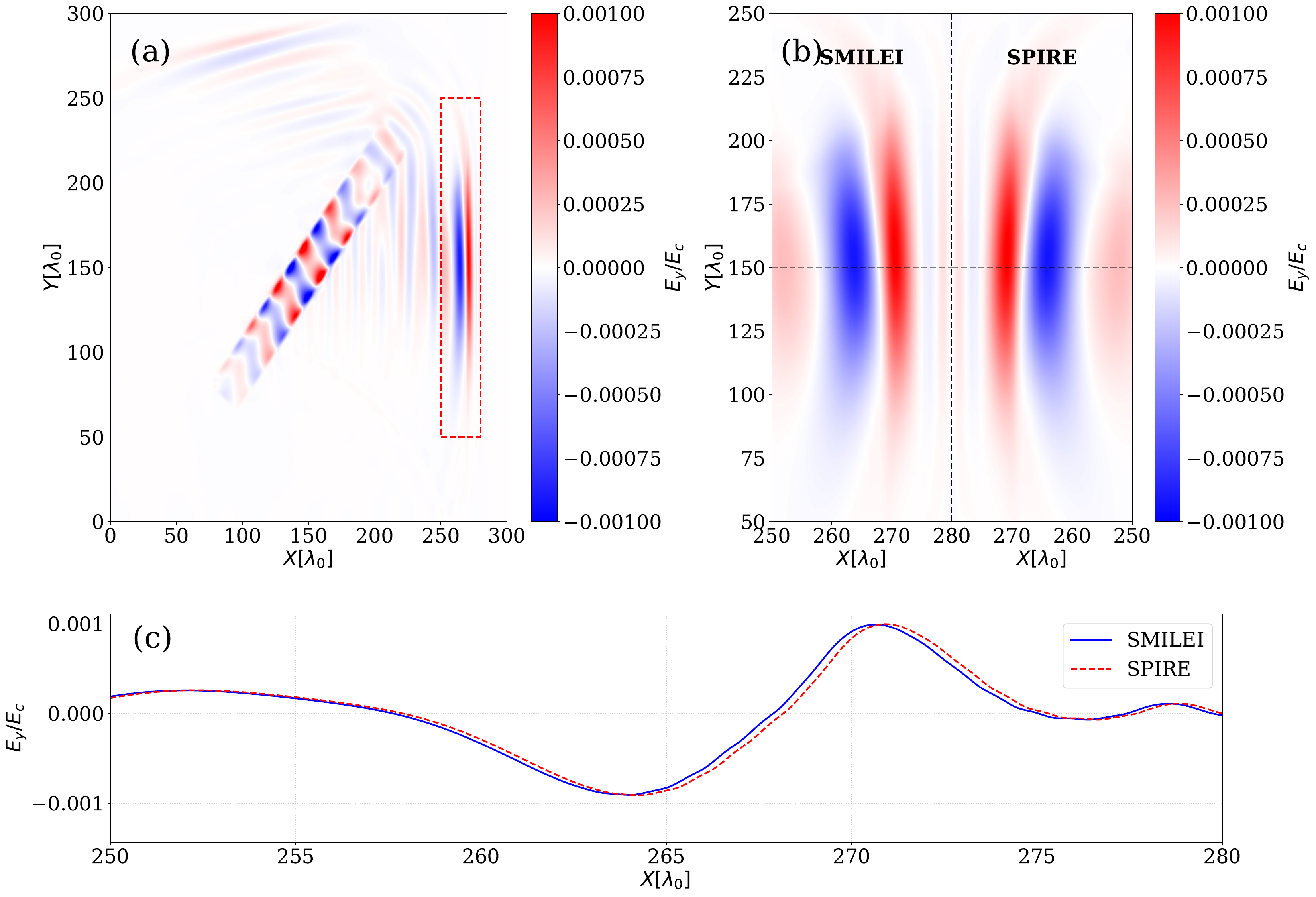}
    \caption{2D benchmark results for laser-interaction with a thin plasma layer: (a) $E_y$ field distribution at the 300th cycle; (b) Comparison of local field distributions between SPIRE and SMILEI; (c) Line-out comparison of the $E_y$ field at $y=150\lambda$.}
    \label{fig:2d_benchmark_subfigure}
\end{figure}

Figure~\ref{fig:2d_benchmark_subfigure} shows the $E_y$ distribution at the 300th cycle after a Gaussian filter has been applied to suppress high-frequency components. The result is a single-cycle electromagnetic wave with a wavelength of approximately $17\text{--}18\lambda$, corresponding to a frequency of 16.7~THz. This value agrees with the Gaussian-unit plasma frequency $\omega_p=\sqrt{4\pi n e^2/m}\approx15$~THz. The local field distributions and the lineout at $y=150\lambda$ also agree closely between SPIRE and SMILEI, supporting the accuracy of the simulation framework.

\subsubsection{Moving window: Laser Wakefield Acceleration}

In laser wakefield acceleration (LWFA), an intense laser pulse propagating
through an underdense plasma drives a relativistic plasma wave through its
ponderomotive force, producing longitudinal electric fields capable of
accelerating charged particles~\cite{TajimaDawson1979}. In the nonlinear
regime, the laser pulse expels plasma electrons from the propagation axis
and forms an electron-depleted cavity, commonly referred to as a plasma
bubble~\cite{Pukhov2002LWFA}. Electrons injected into an appropriate phase
of the wake can be trapped and accelerated to high energies over relatively
short distances, as demonstrated in the first experiments producing
quasi-monoenergetic electron bunches
\cite{Mangles2004LWFA,Geddes2004LWFA,Faure2004LWFA}.
2D PIC simulations were performed in a 
Cartesian $x$--$y$ geometry, with the laser propagating 
along the $+x$ direction. Length and time were normalized 
to the laser wavelength $\lambda_0 = 1 \rm{\mu m}$ and optical period 
$T_0=2\pi/\omega_0$, respectively.

The simulation window had dimensions $L_x\times L_y=60\lambda_0\times40\lambda_0$ 
and was discretized using 24 cells per laser wavelength in both directions, 
corresponding to $\Delta x=\Delta y=\lambda_0/24$. The time step was $\Delta t=T_0/48$, 
satisfying the electromagnetic Courant stability condition, and the 
total simulation duration was $160T_0$.

A linearly $s$-polarized Gaussian laser pulse with a normalized vector potential $a_0=2.0$
was injected from the left boundary. The pulse had a temporal full width at half maximum of $5T_0$ and a transverse waist of $w_0=5\lambda_0$. It was focused at $x=0$ and centered transversely at $y=20\lambda_0$, with an initial carrier phase of zero.

The initially cold and charge-neutral plasma consisted of electrons and singly charged ions with a mass ratio of $m_i/m_e=1836$. The plasma density increased linearly from zero at $x=2\lambda_0$ to a plateau value of $n_0=4.94\times10^{-3}n_c$ over a $2\lambda_0$ entrance ramp. This was followed by a $160\lambda_0$ uniform-density section and a $2\lambda_0$ exit ramp. Both species were initialized with four macroparticles per cell and zero initial momentum. A moving window propagating at the speed of light, $v_{\mathrm{w}}=c$, was activated at $t=30T_0$, with fresh charge-neutral plasma injected through the advancing boundary. Non-periodic boundary conditions were employed in both spatial directions.

\begin{figure}[htbp]
    \centering
    \includegraphics[width=\columnwidth]{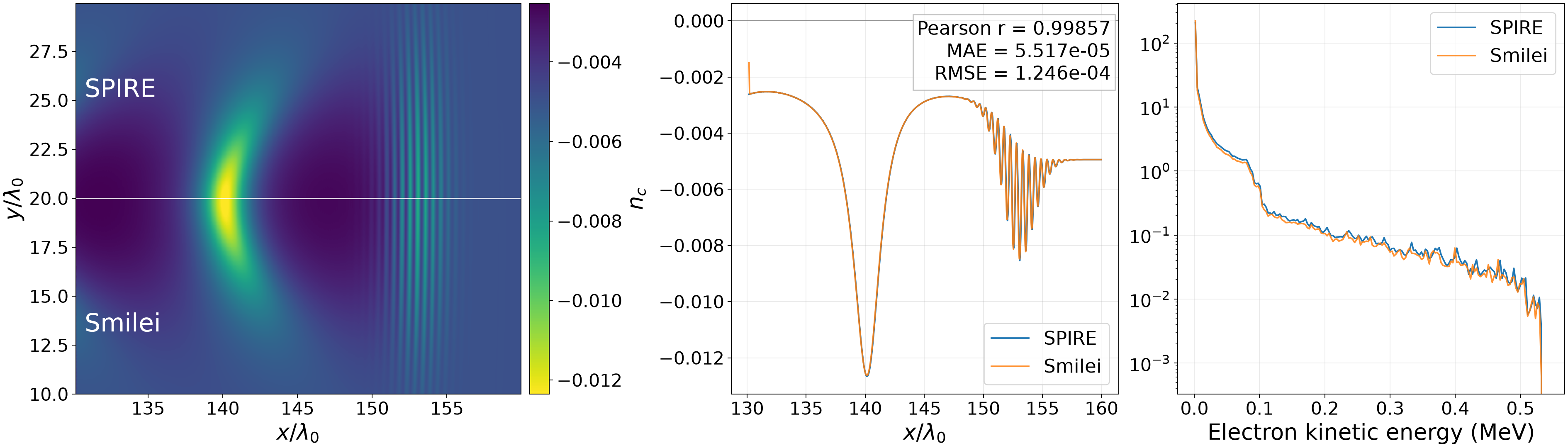}
    \caption{Comparison of results for the laser wakefield acceleration benchmark: 
    (a) Comparison of the electron density $n_e$; 
    (b) 1D distribution of $n_e$ along the axis, where the blue solid and orange dashed lines represent the results from SPIRE and SMILEI, respectively.
    (c) Comparison of the electron energy spectra.}
    \label{fig:moving_window}
\end{figure}

\begin{figure}[htbp]
    \centering
    \includegraphics[width=\columnwidth]{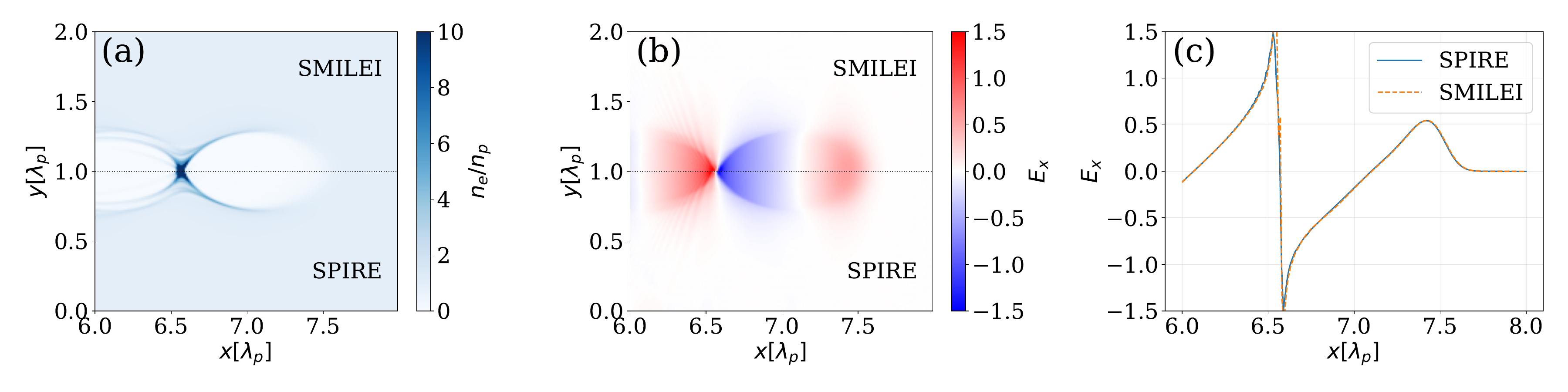}
    \caption{Comparison of results for the 3D high-energy electron-beam-driven wakefield acceleration benchmark: (a) 2D slice comparison of the electron density $n_e/n_p$; (b) 2D slice comparison of the longitudinal electric field $E_x$. In both (a) and (b), the upper and lower halves correspond to the SMILEI and SPIRE results, respectively. (c) 1D distribution of $E_x$ along the axis, where the blue solid and orange dashed lines represent the results from SPIRE and SMILEI, respectively.}
    \label{fig:3d_benchmark_electron}
\end{figure}

\begin{figure}[htbp]
    \centering
    \includegraphics[width=\columnwidth]{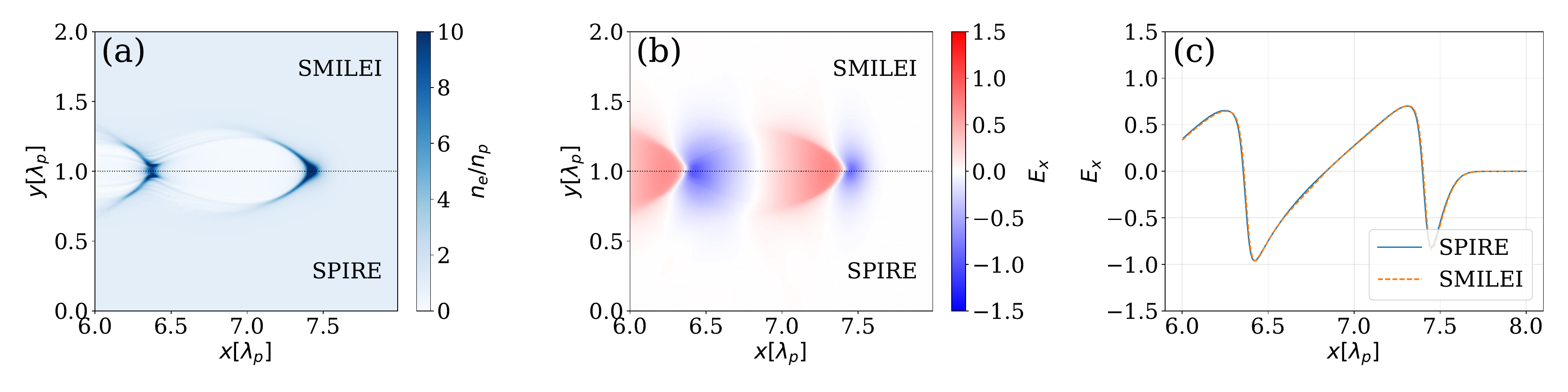}
    \caption{Comparison of results for the 3D high-energy proton-beam-driven wakefield acceleration benchmark: (a) 2D slice comparison of the electron density $n_e/n_p$; (b) 2D slice comparison of the longitudinal electric field $E_x$. In both (a) and (b), the upper and lower halves correspond to the SMILEI and SPIRE results, respectively. (c) 1D distribution of $E_x$ along the axis, where the blue solid and orange dashed lines represent the results from SPIRE and SMILEI, respectively.}
    \label{fig:3d_benchmark_proton}
\end{figure}

\begin{figure*}[htbp]
    \centering
    \includegraphics[width=1.0\textwidth]{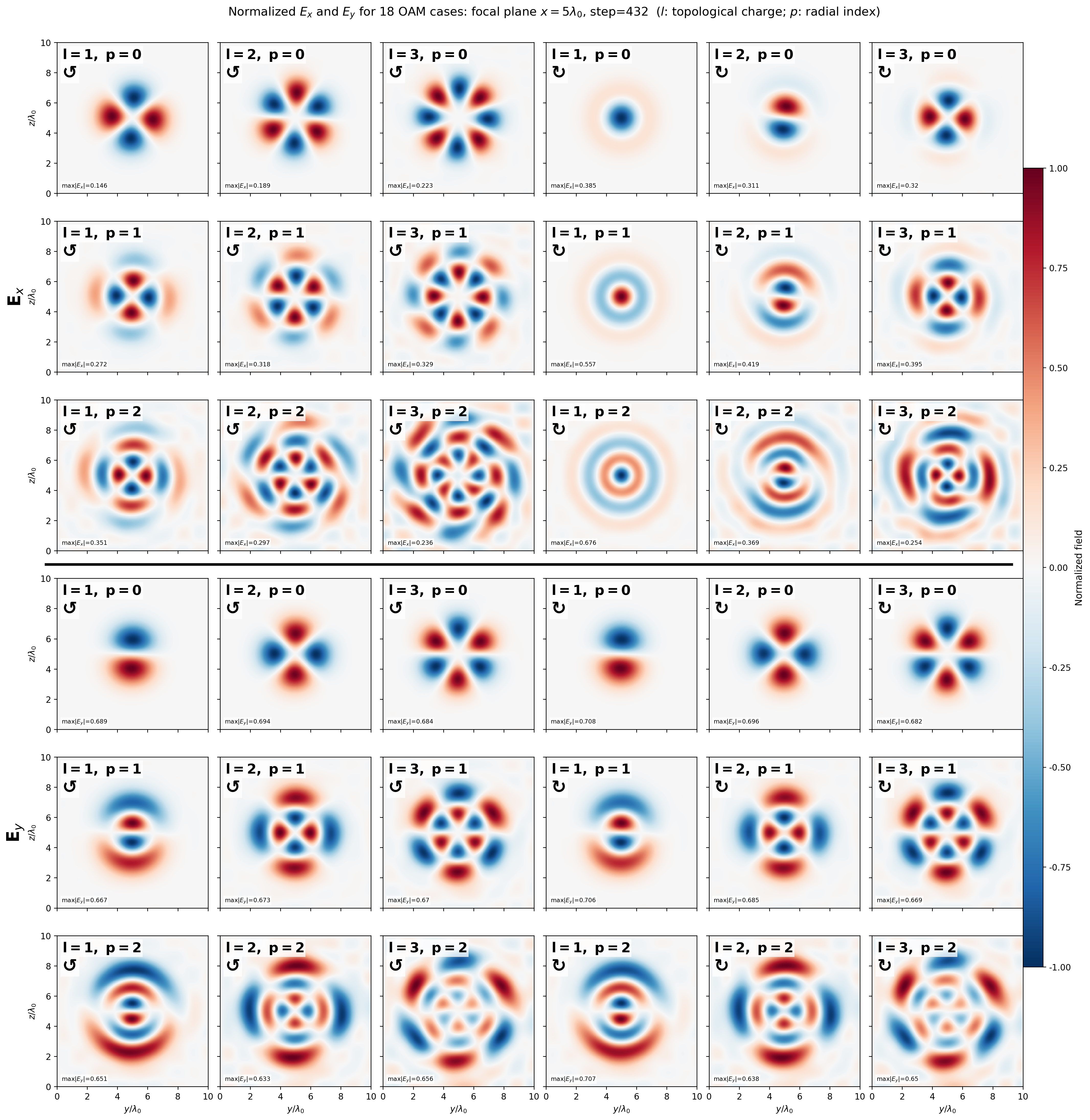}
    \caption{
Normalized transverse distributions of \(E_x\) and \(E_y\) for the 18 OAM
configurations at the focal plane \(x=5\lambda_0\) and at time
\(t=9T_0\) (time step 432). The upper three rows show \(E_x\), whereas the
lower three rows show \(E_y\). 
}
\label{fig:oam18_exey}
\end{figure*}

\subsubsection{3D Case: High-Energy Particle Beam Driven Wakefield Acceleration}

When a high-energy, high-density charged-particle bunch propagates
through a plasma at a relativistic velocity, its collective
electromagnetic field perturbs the background plasma electrons and
excites a trailing plasma wakefield
\cite{Chen1985PWFA,Caldwell2009ProtonPWFA}. The plasma response depends
on the sign of the driver charge. A sufficiently dense electron driver
repels the plasma electrons and, in the nonlinear blowout regime,
produces an electron-depleted ion cavity surrounded by a thin electron
sheath \cite{Lu2006Blowout}. By contrast, a proton driver attracts the
plasma electrons toward the propagation axis and excites a wake through
the resulting collective plasma oscillation. In either case, the wake
can support longitudinal accelerating fields at the GV/m scale or
above, substantially exceeding those available in conventional
radio-frequency accelerators
\cite{Blumenfeld2007EnergyDoubling,Adli2018AWAKE}. A properly phased
witness bunch can therefore be trapped and accelerated by the
longitudinal wakefield.

In this benchmark, either a 1-GeV electron bunch or a 1-GeV proton
bunch is used as the driver. The driver is initialized with a
bi-Gaussian density distribution, with transverse and longitudinal
sizes of $k_p\sigma_r=0.5$ and $k_p\sigma_x=0.5$, respectively, and a
peak density ratio of $n_b/n_p=4$. The plasma occupies
$2\lambda<x<8\lambda$, $0<y<2\lambda$, and $0<z<2\lambda$. The
$8\lambda\times2\lambda\times2\lambda$ simulation domain is resolved
by $800\times200\times200$ cells. Before time integration, the initial
electromagnetic fields of the relativistic driver bunch are obtained
self-consistently by iteratively solving Poisson's equation in the
beam rest frame and subsequently applying a Lorentz transformation to
the laboratory frame~\cite{MASSIMO2016378}. Figures~
\ref{fig:3d_benchmark_electron} and~
\ref{fig:3d_benchmark_proton} compare the three-dimensional
wakefields driven by the electron and proton bunches, respectively.

\subsubsection{LG Laser Injection}

A systematic parameter scan consisting of 18 vacuum simulations was performed
to investigate the transverse electric-field structures of tightly focused
optical angular momentum (OAM) modes. The simulations cover all combinations
of the radial index \(p\in\{0,1,2\}\), the topological charge
\(\ell\in\{1,2,3\}\), and two opposite circular-polarization states. Except for
\(p\), \(\ell\), and the polarization handedness, all simulations employ identical
physical and numerical parameters. The computational domain has dimensions
\(L_x\times L_y\times L_z=10\lambda_0\times10\lambda_0\times10\lambda_0\),
with a uniform spatial resolution
\(\Delta x=\Delta y=\Delta z=\lambda_0/24\) and a temporal resolution
\(\Delta t=T_0/48\). A vectorial OAM pulse with normalized amplitude
\(a_0=1\), waist \(w_0=\lambda_0\), and duration \(10T_0\) is injected at
\(x=\lambda_0\) and focused at
\((x_f,y_f,z_f)=(5,5,5)\lambda_0\). The numerical aperture and pupil filling
factor are set to \(0.6\) and \(0.4\), respectively, while the carrier phase is
zero. The simulations are performed in vacuum without charged particles,
radiation reaction, pair creation, or other QED processes. Periodic boundary
conditions and the moving window are disabled. Each calculation uses an
\(8\times8\times8\) MPI decomposition and is evolved for \(10T_0\).

Figure~\ref{fig:oam18_exey} compares the focal-plane \(E_x\) and \(E_y\)
distributions for all 18 OAM configurations. Increasing the topological charge
primarily increases the azimuthal complexity of the field pattern, whereas
increasing the radial index introduces additional radial nodes and concentric
field structures. Reversing the polarization handedness substantially modifies
the longitudinally induced \(E_x\) component, while the dominant transverse
component \(E_y\) retains a comparatively similar modal structure for the two
opposite handedness states.

\section{Physical Models for Strong-Field QED and Spin Dynamics}

In the regime of ultra-intense laser fields, where the normalized laser 
amplitude $a_0 \gg 1$, the formation lengths for photon emission and 
electron-positron pair production are significantly smaller than both 
the laser wavelength and the characteristic scales of electronic/positronic 
trajectories. This hierarchy of scales justifies the application of 
the local constant field approximation (LCFA) for modeling these 
interactions \cite{Ritus1985,Baier1998,Khokonov2010}. Here, $E_0$ and $\omega_0$ 
denote the laser field amplitude and frequency, respectively, while $e$ and $m$ 
represent the electron charge and mass. We adopt relativistic units ($\hbar = c = 1$) 
throughout this section. The stochastic quantum processes of photon emission and pair 
production are treated using QED methodologies~\cite{Elkina2011,RIDGERS2014273,PhysRevE.92.023305}, where the transition probabilities 
are determined by the local particle trajectories. Specifically, these probabilities 
depend on the local values of the dimensionless quantum nonlinearity parameters 
$\chi_{e,\gamma} \equiv |e| \sqrt{-(F_{\mu\nu}p^\nu)^2}/{m_e}^3$, 
where $F_{\mu\nu}$ is the electromagnetic field tensor and $p^\nu$ is the 
four-momentum of the electron (or photon).

The quantum transition probabilities implemented in this framework are derived 
via the Baier-Katkov operator method within the LCFA limit \cite{Baier1998}. 
The photon emission probability is resolved with respect to the spin states 
of the electron (or positron) before and after emission, as well as the polarization 
state of the emitted photon. Similarly, the pair production probability depends 
on the polarization of the parent photon and the spin states of the secondary 
electron-positron pair. The energy and polarization of the emitted photons and 
secondary particles are determined by sampling the theoretical spectral 
probabilities using Monte Carlo methods. In the relativistic regime, the 
emission is assumed to be collinear with the momentum of the parent particle. 
In every time step, the electron (or positron) polarization vector 
$\boldsymbol{S}_i$ undergoes a transition to $\boldsymbol{S}_f^R$ 
following a photon emission event. Here, the 3D vectors $\boldsymbol{S}_i$ 
and $\boldsymbol{S}_f^R$ represent the spin polarization vectors in the 
particle's rest frame for the initial and final mixed states, respectively, 
which correspond to the average polarization resulting from the scattering 
process \cite{Berestetskii1982}.

If no photon is emitted during a given time step, 
the electron spin still undergoes no-emission spin 
evolution to a state $\boldsymbol{S}_f^{NR}$ owing to one-loop corrections 
to the electron mass operator~\cite{Meuren2011,Ilderton2020,Torgrimsson_2021}. 
Between discrete quantum events, spin precession is governed by 
the Thomas--Bargmann--Michel--Telegdi (T-BMT) equation~\cite{Bargmann1959}, 
including the field-dependent anomalous magnetic moment (AMM) arising from 
vertex corrections~\cite{Baier1972}. The classical orbital motion is simultaneously 
governed by the Lorentz equation. The framework also incorporates macroscopic QED 
vacuum effects, namely vacuum birefringence (VB) and vacuum dichroism (VD), 
which arise because radiative corrections involving virtual particles modify 
photon propagation in strong background fields.

\subsection{Algorithm for Photon Emission Events}
In this work, the photon-emission probability is formulated as follows~\cite{Liyf2020,Song2021BreitWheeler}:

\begin{equation}
\frac{\mathrm{d}^2 W_{r a d}}{\mathrm{~d} u \mathrm{~d} t}=\frac{C_R}{4}\left(F_0+\boldsymbol{\xi} \cdot \mathbf{F}\right),
\label{equation:S1}
\end{equation}

where $C_R = \alpha m/[\sqrt{3}\pi \gamma_e(1 + u)^3]$, $\alpha$ is the fine-structure constant, $\gamma_e$ is the electron Lorentz factor, $u = \omega_\gamma/(\varepsilon_i - \omega_\gamma)$, $\omega_\gamma$ is the emitted-photon energy, and $\varepsilon_i$ is the initial electron energy. The auxiliary quantities $F_0$ and $\mathbf F$ in Eq.~\ref{equation:S1} are defined as follows:

\begingroup
\allowdisplaybreaks[3]
\begin{align}
F_0={}&-(1+u)\operatorname{IntK}_{\frac13}(u')
+(u^2+2u+2)\mathrm K_{\frac23}(u')\notag\\*
&-u(\mathbf S_i\cdot\hat{\mathbf e}_2)\mathrm K_{\frac13}(u')\notag\\*
&-\mathbf S_f\Bigl\{(1+u)
[\operatorname{IntK}_{\frac13}(u')-2\mathrm K_{\frac23}(u')]\mathbf S_i\notag\\*
&\hspace{5em}+u(1+u)\mathrm K_{\frac13}(u')\hat{\mathbf e}_2\notag\\*
&\hspace{5em}+u^2[\operatorname{IntK}_{\frac13}(u')
-\mathrm K_{\frac23}(u')]
(\mathbf S_i\cdot\hat{\mathbf e}_v)\hat{\mathbf e}_v\Bigr\}
\notag\displaybreak[3]\\
F_1={}&(\mathbf S_i\cdot\hat{\mathbf e}_1)u(1+u)\mathrm K_{\frac13}(u')\notag\\*
&+(\mathbf S_f\cdot\hat{\mathbf e}_1)u\mathrm K_{\frac13}(u')\notag\\*
&+(\mathbf S_f\times\mathbf S_i)\hat{\mathbf e}_v
\left(\frac{u^2}{2}+u\right)\mathrm K_{\frac23}(u')\notag\\*
&-\frac{u^2}{2}\Bigl[(\mathbf S_i\cdot\hat{\mathbf e}_1)
(\mathbf S_f\cdot\hat{\mathbf e}_2)\notag\\*
&\hspace{6em}+(\mathbf S_f\cdot\hat{\mathbf e}_1)
(\mathbf S_i\cdot\hat{\mathbf e}_2)\Bigr]
\operatorname{IntK}_{\frac13}(u')
\notag\displaybreak[3]\\
F_2={}&-(\mathbf S_i\cdot\hat{\mathbf e}_v)u
\operatorname{IntK}_{\frac13}(u')\notag\\*
&-(\mathbf S_f\cdot\hat{\mathbf e}_v)u(1+u)
\operatorname{IntK}_{\frac13}(u')\notag\\*
&+[(\mathbf S_i\cdot\hat{\mathbf e}_v)
+(\mathbf S_f\cdot\hat{\mathbf e}_v)](2+u)u\mathrm K_{\frac23}(u')\notag\\*
&-(\mathbf S_f\times\mathbf S_i)\hat{\mathbf e}_1
\left(\frac{u^2}{2}+u\right)\mathrm K_{\frac13}(u')\notag\\*
&-\frac{u^2}{2}\Bigl[(\mathbf S_i\cdot\hat{\mathbf e}_v)
(\mathbf S_f\cdot\hat{\mathbf e}_2)\notag\\*
&\hspace{6em}+(\mathbf S_f\cdot\hat{\mathbf e}_v)
(\mathbf S_i\cdot\hat{\mathbf e}_2)\Bigr]\mathrm K_{\frac13}(u')
\notag\displaybreak[3]\\
F_3={}&(1+u)\mathrm K_{\frac23}(u')
-u(1+u)(\mathbf S_i\cdot\hat{\mathbf e}_2)\mathrm K_{\frac13}(u')\notag\\*
&+(\mathbf S_i\cdot\mathbf S_f)
\left(\frac{u^2}{2}+u+1\right)\mathrm K_{\frac23}(u')\notag\\*
&-u(\mathbf S_f\cdot\hat{\mathbf e}_2)\mathrm K_{\frac13}(u')\notag\\*
&-\frac{u^2}{2}\Bigl\{(\mathbf S_i\cdot\hat{\mathbf e}_v)
(\mathbf S_f\cdot\hat{\mathbf e}_v)\mathrm K_{\frac23}(u')\notag\\*
&\hspace{3em}+[(\mathbf S_i\cdot\hat{\mathbf e}_1)
(\mathbf S_f\cdot\hat{\mathbf e}_1)\notag\\*
&\hspace{6em}-(\mathbf S_i\cdot\hat{\mathbf e}_2)
(\mathbf S_f\cdot\hat{\mathbf e}_2)]
\operatorname{IntK}_{\frac13}(u')\Bigr\}.
\label{equation:S2}
\end{align}
\endgroup
where $u' = 2u/(3\chi_e)$ and $\operatorname{IntK}_{1/3}(u') \equiv \int_{u'}^\infty \mathrm{d}z \, K_{1/3}(z)$; $K_n$ denotes the $n$-th order modified Bessel function of the second kind. The unit vector $\hat{\mathbf{e}}_1$ is oriented along the transverse component of the electron acceleration, and $\hat{\mathbf{e}}_2 = \hat{\mathbf{e}}_v \times \hat{\mathbf{e}}_1$ (where $\hat{\mathbf{e}}_v$ is the unit vector along the electron velocity). The vectors $\mathbf{S}_i$ and $\mathbf{S}_f$ represent the electron spin polarization before and after radiation, respectively, with $|\mathbf{S}_{i,f}| \leq 1$.

Within a time step $\Delta t$, the probability that an electron emits a photon of energy $\omega_\gamma = \delta \varepsilon_i$ ($0 < \delta < 1$) is calculated from Eq.~\ref{equation:S1} as $W_{\text{rad}}(\delta) = \sum_{S_i,S_f,\xi} [\mathrm{d}W_{\text{rad}}/\mathrm{d}u][\mathrm{d}u/\mathrm{d}\delta]\Delta t$, where $\delta=r_1^3$ and $r_1$ is uniformly distributed on $[0,1]$. A second random number $r_2\in[0,1]$ determines whether emission occurs. For ultrarelativistic electrons, the characteristic emission angle is of order $1/\gamma_e$, so the emitted photon is assumed to propagate along the electron velocity. The approximations used here retain the leading order in $1/\gamma_e$; further details are given in Ref.~\cite{Gonoskov2015}.

For a given emitted-photon energy, the final electron spin state is determined from Eq.~\ref{equation:S1} after summing over photon polarization states:

\begin{equation}
\begin{aligned}
\frac{\mathrm{d}^2 \bar{W}_{r a d}}{\mathrm{~d} u \mathrm{~d} t} & =\frac{C_R}{2} F_0, \\
& =F_{00}+\mathbf{S}_f \cdot \mathbf{F}_{01},
\end{aligned}
\label{equation:S3}
\end{equation}

where

\begin{equation}
\begin{aligned}
F_{00}= & -(1+u) \operatorname{IntK}_{\frac{1}{3}}\left(u^{\prime}\right)+\left(u^2+2 u+2\right) \mathrm{K}_{\frac{2}{3}}\left(u^{\prime}\right) \\
& -\mathbf{S}_i \cdot \hat{\mathbf{e}}_2 u \mathrm{~K}_{\frac{1}{3}}\left(u^{\prime}\right), \\
\mathbf{F}_{01}= & -(1+u)\left[\operatorname{IntK}_{\frac{1}{3}}\left(u^{\prime}\right)-2 \mathrm{~K}_{\frac{2}{3}}\left(u^{\prime}\right)\right] \mathbf{S}_i \\
& -u(1+u) \mathrm{K}_{\frac{1}{3}}\left(u^{\prime}\right) \hat{\mathbf{e}}_2 \\
& -u^2\left[\operatorname{IntK}_{\frac{1}{3}}\left(u^{\prime}\right)-\mathrm{K}_{\frac{2}{3}}\left(u^{\prime}\right)\right]\left(\mathbf{S}_i \cdot \hat{\mathbf{e}}_v\right) \hat{\mathbf{e}}_v,
\end{aligned}
\label{equation:S4}
\end{equation}

Following the emission of a photon, the electron spin undergoes a transition to $\mathbf{S}_f^R$:

\begin{equation}
\mathbf{S}_f^R=\frac{\mathbf{F}_{01}}{F_{00}} .
\label{equation:S5}
\end{equation}

If the photon-emission event is rejected, the electron spin still undergoes no-emission spin evolution:

\begin{equation}
\frac{\mathrm{d}^2 \bar{W}_{\mathrm{no\text{-}rad}}}{\mathrm{d}u\,\mathrm{d}t}
=F_{00}^{\mathrm{no}}+\mathbf{S}_f^{NR}\cdot\mathbf{F}_{01}^{\mathrm{no}}
\label{equation:S6}
\end{equation}

\begin{equation}
\mathbf{S}_f^{NR}=\frac{\mathbf{F}_{01}^{\mathrm{no}}}{F_{00}^{\mathrm{no}}} .
\label{equation:S7}
\end{equation}

Here,
\begin{equation}
\begin{aligned}
F_{00}^{\mathrm{no}}={}&1-\int C_R\Bigl[
-(1+u)\operatorname{IntK}_{\frac{1}{3}}(u')\\
&\quad +(u^2+2u+2)\mathrm K_{\frac{2}{3}}(u')
-u(\mathbf S_i\cdot\hat{\mathbf e}_2)
\mathrm K_{\frac{1}{3}}(u')\Bigr]\,\mathrm du\,\Delta t,\\
\mathbf F_{01}^{\mathrm{no}}={}&\mathbf S_i\Biggl\{1-\int C_R\Bigl[
-(1+u)\operatorname{IntK}_{\frac{1}{3}}(u')\\
&\quad +(u^2+2u+2)\mathrm K_{\frac{2}{3}}(u')\Bigr]
\,\mathrm du\,\Delta t\Biggr\}\\
&+\hat{\mathbf e}_2\int C_Ru\mathrm K_{\frac{1}{3}}(u')
\,\mathrm du\,\Delta t.
\end{aligned}
\label{equation:S77}
\end{equation}

After photon emission, the Stokes parameters $\xi=(\xi_1,\xi_2,\xi_3)$ of the emitted photon, defined relative to the coordinate axes $\hat{\mathbf{e}}_1$ and $\hat{\mathbf{e}}_2$, are given by:

\begin{equation}
\xi=\frac{\mathbf{F}}{F_0}, \quad \mathbf{F}=\left(F_1, F_2, F_3\right) .
\label{equation:S8}
\end{equation}

\subsection{Algorithm for Electron-Positron Pair Production Events}

Following an approach similar to the algorithm for photon emission, the total pair production probability, obtained by summing over final spin states, is first utilized to determine the occurrence of an electron-positron pair production event. Subsequently, the specific spin states of the pair are assigned based on the corresponding differential probabilities. The pair production probability, which depends on the initial photon polarization, is given by \cite{Baier1998}:

\begin{equation}
\begin{aligned}
\frac{\mathrm d^2\bar W_{\mathrm{pair}}}
{\mathrm d\varepsilon_+\,\mathrm dt}
={}&C_p\Biggl\{|\mathbf e\cdot\hat{\mathbf e}_1|^2\Biggl[
\left(\frac{\omega_\gamma^2}{\varepsilon_+\varepsilon_-}-3\right)
\mathrm K_{\frac{2}{3}}(\rho)
+\operatorname{IntK}_{\frac{1}{3}}(\rho)\Biggr]\\
&\quad+|\mathbf e\cdot\hat{\mathbf e}_2|^2\Biggl[
\left(\frac{\omega_\gamma^2}{\varepsilon_+\varepsilon_-}-1\right)
\mathrm K_{\frac{2}{3}}(\rho)
+\operatorname{IntK}_{\frac{1}{3}}(\rho)\Biggr]\Biggr\}\\
={}&C_p\Biggl\{|\xi|\Biggl[
\left(\frac{\omega_\gamma^2}{\varepsilon_+\varepsilon_-}-2\right)
\mathrm K_{\frac{2}{3}}(\rho)
+\operatorname{IntK}_{\frac{1}{3}}(\rho)\Biggr]\\
&\quad-\xi_3\operatorname{IntK}_{\frac{1}{3}}(\rho)\Biggr\}.
\end{aligned}
\label{equation:S9}
\end{equation}

where $C_p = \frac{\alpha m^2}{\sqrt{3}\pi \omega_\gamma^2}$ and $\mathbf{e}$ denotes the photon polarization state \cite{McMaster1961}; $\omega_\gamma, \varepsilon_+$, and $\varepsilon_-$ represent the energies of the photon, positron, and electron, respectively (satisfying $\omega_\gamma = \varepsilon_+ + \varepsilon_-$); and $\rho = \frac{2\omega_\gamma^2}{3\chi_\gamma \varepsilon_+ \varepsilon_-}$. In this context, the pair production probability can be expressed as:

\begin{equation}
\begin{aligned}
\frac{\mathrm{d}^2 \bar{W}_{\text {pair }}}{\mathrm{d} \varepsilon_{+} \mathrm{d} t}= & C_p\left\{\left[\left(\frac{\omega_\gamma^2}{\varepsilon_{+} \varepsilon_{-}}-2\right) \mathrm{K}_{\frac{2}{3}}(\rho)+\operatorname{IntK}_{\frac{1}{3}}(\rho)\right]\right. \\
& \left.-\xi_3 \operatorname{IntK}_{\frac{1}{3}}(\rho)\right\}
\end{aligned}
\label{equation:S10}
\end{equation}

The spin state of the produced positron is determined by the joint probability distribution involving both the photon polarization and the positron spin:

\begin{equation}
\frac{\mathrm{d}^2 W_{p a i r}}{\mathrm{~d} \varepsilon_{+} \mathrm{d} t}=\frac{C_p}{2}\left(F_0^P+\boldsymbol{\xi} \cdot \mathbf{F}^P\right)
\label{equation:S11}
\end{equation}

where $F_0^P = \Big[ \frac{\omega_\gamma^2}{\varepsilon_+ \varepsilon_-} - 2 \Big] K_{\frac{1}{3}}(\rho) + \text{Int}K_{\frac{1}{3}}(\rho) + \frac{\omega_\gamma}{\varepsilon_-} (\mathbf{S}_+ \cdot \hat{\mathbf{e}}_2) K_{\frac{1}{3}}(\rho)$, $\mathbf{F}^P = (F_1^P, F_2^P, F_3^P)$, and:

\begin{equation}
\begin{aligned}
F_1^P & =\frac{\omega_\gamma}{\varepsilon_{-}}\left(\mathbf{S}_{+} \cdot \hat{\mathbf{e}}_1\right) \mathrm{K}_{\frac{1}{3}}(\rho), \\
F_2^P & =\left[\frac{\omega_\gamma}{\varepsilon_{+}} \operatorname{IntK}_{\frac{1}{3}}(\rho)+\frac{\varepsilon_{+}^2-\varepsilon_{-}^2}{\varepsilon_{+} \varepsilon_{-}} \mathrm{K}_{\frac{2}{3}}(\rho)\right]\left(\mathbf{S}_{+} \cdot \hat{\mathbf{e}}_v\right), \\
F_3^P & =-\mathrm{K}_{\frac{2}{3}}(\rho)-\frac{\omega_\gamma}{\varepsilon_{-}}\left(\mathbf{S}_{+} \cdot \hat{\mathbf{e}}_2\right) \mathrm{K}_{\frac{1}{3}}(\rho),
\end{aligned}
\label{equation:S12}
\end{equation}

Equation~\ref{equation:S11} can be rewritten in a form analogous to Eq.~\ref{equation:S1}:

\begin{equation}
\frac{\mathrm{d}^2 W_{p a i r}}{\mathrm{~d} \varepsilon_{+} \mathrm{d} t}=\frac{C_p}{2}\left(\tilde{F}_0^P+ \mathbf{S_{+}} \cdot \tilde{\mathbf{F}}\right)
\label{equation:S13}
\end{equation}

where $\tilde{F}_0^P = \Big[ \frac{\omega_\gamma^2}{\varepsilon_+ \varepsilon_-} - 2 \Big] K_{\frac{1}{3}}(\rho) + \text{Int}K_{\frac{1}{3}}(\rho) - \xi_3 K_{\frac{2}{3}}(\rho)$, and $\tilde{\mathbf{F}} = \xi_1 \frac{\omega_\gamma}{\varepsilon_-} K_{\frac{1}{3}}(\rho) \hat{\mathbf{e}}_1 + \Big( \frac{\omega_\gamma}{\varepsilon_+} - \xi_3 \frac{\omega_\gamma}{\varepsilon_-} \Big) K_{\frac{1}{3}}(\rho) \hat{\mathbf{e}}_2 + \xi_2 \Bigg[ \frac{\omega_\gamma}{\varepsilon_+} \text{Int}K_{\frac{1}{3}}(\rho) + \frac{\varepsilon_+^2 - \varepsilon_-^2}{\varepsilon_+ \varepsilon_-} K_{\frac{2}{3}}(\rho) \Bigg] \hat{\mathbf{e}}_v$.

The spin state $\mathbf{S}_+$ is thus determined as:

\begin{equation}
\mathbf{S_{+}}=\frac{\tilde{\mathbf{F}}}{\tilde{F}_0^P}
\label{equation:S14}
\end{equation}

The spin state of the electron in the pair can also be determined using the equations above by substituting $(\varepsilon_+, \varepsilon_-, \mathbf{S}_+, \hat{\mathbf{e}}_{1,2})$ with $(\varepsilon_-, \varepsilon_+, \mathbf{S}_-, -\hat{\mathbf{e}}_{1,2})$, respectively.

\subsection{TBMT Equation and Anomalous Magnetic Moment}

In the intervals between photon emission and electron-positron pair production events, the spin precession of an electron (or positron) is governed by the Thomas-Bargmann-Michel-Telegdi (T-BMT) equation \cite{Bargmann1959}:

\begin{equation}
\begin{aligned}
\left(\frac{d \mathbf{S}}{d t}\right)_T= & \mathbf{S} \times \boldsymbol{\Omega} \\
\equiv & \mathbf{S} \times\left[-\left(\frac{g}{2}-1\right) \frac{\gamma_e}{\gamma_e+1}(\boldsymbol{\beta} \cdot \mathbf{B}) \cdot \boldsymbol{\beta}\right. \\
& \left.+\left(\frac{g}{2}-1+\frac{1}{\gamma_e}\right) \mathbf{B}-\left(\frac{g}{2}-\frac{\gamma_e}{\gamma_e+1}\right) \boldsymbol{\beta} \times \mathbf{E}\right],
\end{aligned}
\label{equation:S15}
\end{equation}
where $\mathbf{E}$ and $\mathbf{B}$ are the electric and magnetic fields of the laser, respectively, and $g$ is the gyromagnetic factor of the electron. The expression for $g$ is $g(\chi_e) = 2 + 2\mu(\chi_e)$, where the anomalous magnetic moment is $\mu(\chi_e) = \frac{\alpha}{\pi} \int_0^\infty \frac{y}{(1+y)^3} \mathbf{L}_{\frac{1}{3}}^3 \Big( \frac{2y}{3\chi_e} \Big) \mathrm{d}y$ (and $\mathbf{L}_{\frac{1}{3}}^3(z) = \int_0^\infty \sin\Bigg[ \frac{3z}{2} \Big( x + \frac{x^3}{3} \Big) \Bigg] \mathrm{d}x$). In the regime where $\chi_e \ll 1$, $g$ takes the approximate value of 2.00232.

The T-BMT equation has the same mathematical structure as the rotational part of the Lorentz equation. We therefore use a rotation-only Boris algorithm to integrate the spin precession~\cite{Wan2023}:

\begin{equation}
\begin{gathered}
\mathbf{S}^{\prime}=\mathbf{S}^{n-1 / 2}+\mathbf{S}^{n-1 / 2} \times \mathbf{t}, \\
\mathbf{S}^{n+1 / 2}=\mathbf{S}^{n-1 / 2}+\mathbf{S}^{\prime} \times \mathbf{o}, \\
\mathbf{t}=\frac{q \Delta t}{2} \boldsymbol{\Omega}^n, \\
\mathbf{o}=\frac{2 \mathbf{t}}{1+t^2} .
\end{gathered}
\label{equation:S16}
\end{equation}

\subsection{Vacuum Birefringence and Dichroism}

When an ultra-intense laser field interacts with the vacuum, virtual particle-antiparticle pairs undergo polarization, oriented in a manner analogous to molecules in birefringent materials. Birefringence in a medium results in a polarization-dependent refractive index; similarly, vacuum birefringence (VB) causes the vacuum to exhibit disparate optical properties for different photon polarizations. Dichroism refers to differential absorption based on the polarization state of light. In the context of laser-induced vacuum dichroism (VD), the vacuum selectively absorbs or attenuates photons at different rates depending on their polarization.

Within QED, these processes arise from interference between tree-level and one-loop photon-propagator amplitudes. We describe them using expressions derived from the ``no-pair-production probability''~\cite{Zhuang2023,CAIN,Dai2024,Torgrimsson_2021}.

Following the methodology described in Ref. \cite{Bragin2017}, the treatment of vacuum birefringence is formulated as:

\begin{equation}
\binom{\xi_{f, 1}}{\xi_{f, 2}}=\left(\begin{array}{cc}
\cos \delta \phi & -\sin \delta \phi \\
\sin \delta \phi & \cos \delta \phi
\end{array}\right)\binom{\xi_{i, 1}}{\xi_{i, 2}} .
\label{equation:S17}
\end{equation}

where $\delta\phi = \phi_1 - \phi_2$ is calculated as:

\begin{equation}
\phi_i=-\frac{1}{2 k q} \operatorname{Re}\left[p_i\left(\chi_\gamma\right)\right] \Delta \eta .
\label{equation:S18}
\end{equation}

\begin{equation}
\left[\begin{array}{l}
p_1\left(\eta, \chi_\gamma\right) \\
p_2\left(\eta, \chi_\gamma\right)
\end{array}\right]=\frac{\alpha m^2}{3 \pi} \int_{-1}^{+1} d v\left[\begin{array}{l}
(w-1) \\
(w+2)
\end{array}\right] \frac{f^{\prime}\left(w^{\prime}\right)}{w^{\prime}},
\label{equation:S19}
\end{equation}

where $w = 4/(1-v^2)$, $w' = [w/\chi_\gamma]^{1/3}$, $f(w') = \pi [\operatorname{Gi}(w') + i\text{Ai}(w')]$, and $\eta = \omega_0 t - k_0 z$ represents the laser phase.

Vacuum dichroism is implemented through the "no-pair-production evolution," which arises from an intrinsic physical process: photons in specific polarization states are absorbed with varying probabilities. This polarization-dependent absorption leads to the evolution of the polarization state of the surviving photons, a process described by the no-pair-production probability according to Refs. \cite{Zhuang2023,Dai2024}:

\begin{equation}
W_{\mathrm{noP}}=F_0^L+\boldsymbol{\xi}_f \cdot \mathbf{F}^L
\label{equation:S20}
\end{equation}

Specifically,

\begin{widetext}
\begin{align}
\boldsymbol{\xi}_f & =\frac{\boldsymbol{\xi}_i\left\{1-\int_0^1 d x C_p\left[\operatorname{IntK}_{\frac{1}{3}}(\rho)+\left(\frac{x}{1-x}+\frac{1-x}{x}\right) \mathrm{K}_{\frac{2}{3}}(\rho)\right] \Delta t\right\}+\int_0^1 d x C_p \hat{\mathbf{e}}_3^{\prime} \mathrm{K}_{\frac{2}{3}}(\rho) \Delta t}{1-\int_0^1 d x C_p\left[\left(\frac{x}{1-x}+\frac{1-x}{x}\right) \mathrm{K}_{\frac{2}{3}}(\rho)+\operatorname{IntK}_{\frac{1}{3}}(\rho)-\xi_3 \mathrm{~K}_{\frac{2}{3}}(\rho)\right] \Delta t} \nonumber \\
& \approx \boldsymbol{\xi}_i\left\{1+\int_0^1 d x C_p\left[\left(\frac{x}{1-x}+\frac{1-x}{x}\right) \mathrm{K}_{\frac{2}{3}}(\rho)+\operatorname{IntK}_{\frac{1}{3}}(\rho)-\xi_3 \mathrm{~K}_{\frac{2}{3}}(\rho)\right] \Delta t\right\} \nonumber \\
& +\left\{\int_0^1 d x C_p \hat{\mathbf{e}}_3^{\prime} \mathrm{K}_{\frac{2}{3}}(\rho)-\boldsymbol{\xi}_i \int_0^1 d x C_p\left[\operatorname{IntK}_{\frac{1}{3}}(\rho)+\left(\frac{x}{1-x}+\frac{1-x}{x}\right) \mathrm{K}_{\frac{2}{3}}(\rho)\right]\right\} \Delta t \nonumber \\
& =\boldsymbol{\xi}_i+\int_0^1 d x C_p \mathrm{~K}_{\frac{2}{3}}(\rho)\left[\hat{\mathbf{e}}_3^{\prime}-\boldsymbol{\xi}_i \xi_3\right] \Delta t
\label{equation:S21}
\end{align}
\end{widetext}

\begin{figure}[htbp]
    \centering
    \includegraphics[width=\columnwidth]{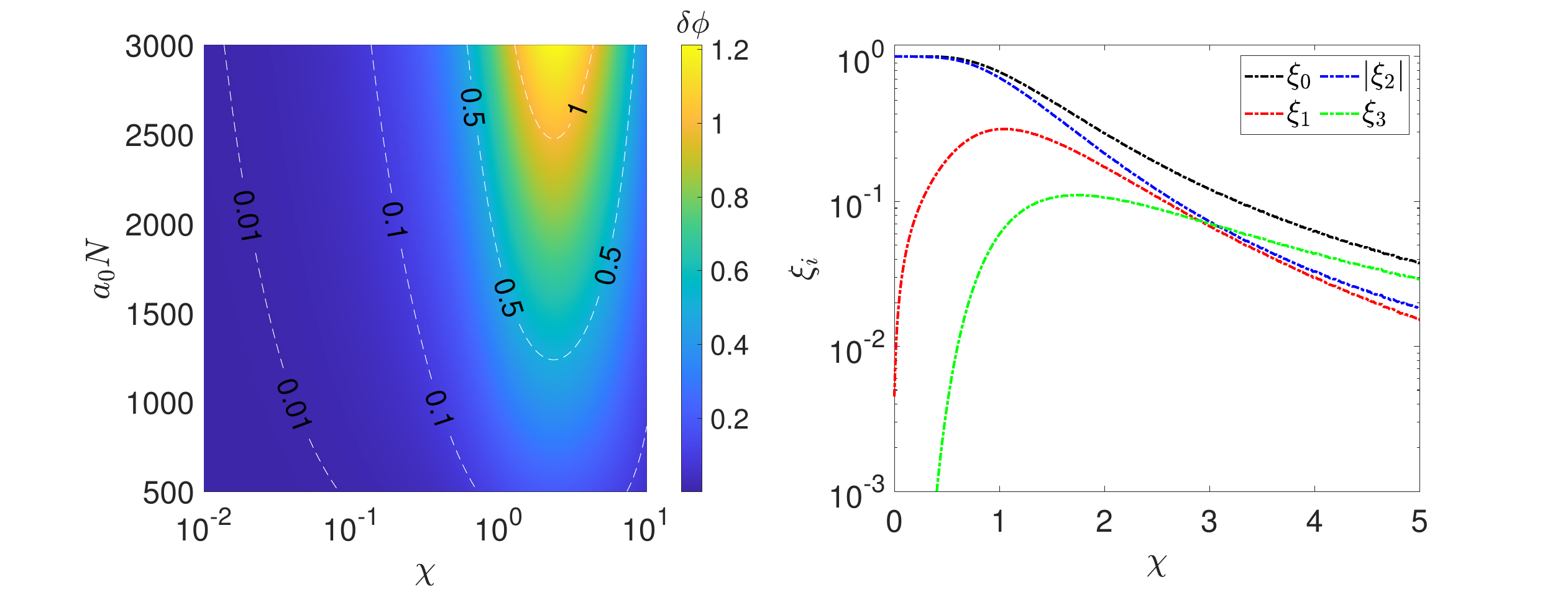}
    \caption{Comparison with the results in Figs.~4 and 5 of Ref.~\cite{Bragin2017}. (a) Phase difference $\delta\phi$ as a function of $\chi$ and $a_0N$ for a rectangular pulse containing $N$ laser cycles. (b) Final Stokes parameters of a gamma-ray photon with initial Stokes vector $\boldsymbol{\xi}=(1,0,-1,0)$ after propagation through an ELI-NP 10-PW laser pulse.}
    \label{figure:vpbenchmark}
\end{figure}

We compare the present results with those of Ref.~\cite{Bragin2017}; see Fig.~\ref{figure:vpbenchmark}. That work used the polarization-operator method in a nonperturbative framework to calculate the photon polarization across the beam while accounting for both vacuum birefringence and dichroism in a strong external laser field.

The no-pair-production evolution used here is analogous to the no-emission spin evolution of electrons because both are governed by polarization-dependent survival probabilities. These effects arise from interference between tree-level and one-loop amplitudes. Following Ref.~\cite{Torgrimsson_2021}, the QED loop contributions are evaluated within the local constant field approximation. The leading-order contribution is

\begin{equation}
\mathbb{P}^0=\frac{1}{2}\left(1+\boldsymbol{\xi}^i \cdot \boldsymbol{\xi}^f\right)
\label{equation:S22}
\end{equation}

where $\boldsymbol{\xi}^i$ and $\boldsymbol{\xi}^f$ represent the initial and final states of the photon Stokes vector, respectively. The $\mathcal{O}(\alpha)$ loop contribution, including the interference diagrams, is given by:

\begin{equation}
\mathbb{P}^{\mathrm{L}}=\left\langle\mathbb{P}^{\mathrm{L}}\right\rangle+\mathbf{P}_0^{\mathrm{L}} \cdot \mathbf{n}_0+\mathbf{P}_1^{\mathrm{L}} \cdot \mathbf{n}_1+\mathbf{n}_1 \cdot \mathbf{P}_{10}^{\mathrm{L}} \cdot \mathbf{n}_0,
\label{equation:S23}
\end{equation}

where $\mathbf{n}_0 = \boldsymbol{\xi}^i$ and $\mathbf{n}_1 = \boldsymbol{\xi}^f$. For a linearly polarized laser field, the components of Eq.~\ref{equation:S23} are

\begin{equation}
\left\{\left\langle\mathbb{P}^{\mathrm{L}}\right\rangle, \mathbf{P}_0^{\mathrm{L}}, \ldots\right\}=\frac{\alpha}{2} \int \frac{\mathrm{~d} \sigma}{b_0} \int_0^1 \mathrm{~d} s\left\{\left\langle\hat{\mathbb{R}}^{\mathrm{L}}\right\rangle, \hat{\mathbf{R}}_0^{\mathrm{L}}, \ldots\right\},
\label{equation:S24}
\end{equation}

\begin{equation}
\left\langle\hat{\mathbb{R}}^{\mathrm{L}}\right\rangle=-\mathrm{Ai}_1(\xi)+\kappa \frac{\mathrm{Ai}^{\prime}(\xi)}{\xi}
\label{equation:S25}
\end{equation}

\begin{equation}
\hat{\mathbf{R}}_0^{\mathrm{L}}=\hat{\mathbf{R}}_1^{\mathrm{L}}=-\frac{\operatorname{Ai}^{\prime}(\xi)}{\xi} \boldsymbol{\epsilon}_{\mathrm{E}},
\label{equation:S26}
\end{equation}

\begin{equation}
\hat{\mathbf{R}}_{10, i j}^{\mathrm{L}}=\left\langle\hat{\mathbb{R}}^{\mathrm{L}}\right\rangle \mathbf{1}+\frac{\operatorname{Gi}^{\prime}(\xi)}{\xi} \varepsilon_{3 i j},
\label{equation:S27}
\end{equation}

Here, $s=\varepsilon/\omega$, where $\varepsilon$ and $\varepsilon_+$ are the energies of the produced electron and positron, respectively, and $\omega$ is the energy of the parent photon. The parameter $\xi=(r/\chi_\gamma)^{2/3}$ is defined in terms of the photon quantum parameter $\chi_\gamma\equiv |e|\sqrt{-(F_{\mu\nu}k^\nu)^2}/m^3$. The quantity $b_0=kq$ is the dimensionless longitudinal momentum of the incident photon. Furthermore, $\mathrm{Ai}_1(\xi)=\int_{\xi}^{\infty}\mathrm{d}t\,\mathrm{Ai}(t)$, $\kappa=s/(1-s)+(1-s)/s$, $r=1/s+1/(1-s)$, $\boldsymbol{\epsilon}_{\mathrm E}=\{0,0,1\}$, and $\varepsilon_{3ij}$ is the Levi--Civita tensor, with $\varepsilon_{123}=1$. The functions $\mathrm{Ai}(x)$ and $\operatorname{Gi}(x)$ denote the Airy function and the Scorer function, respectively.

\begin{equation}
\operatorname{Ai}(x)=\frac{1}{\pi} \sqrt{\frac{x}{3}} K_{1 / 3}\left(\frac{2}{3} x^{3 / 2}\right)
\label{equation:S28}
\end{equation}

\begin{equation}
\operatorname{Ai}^{\prime}(x)=-\frac{x^{1 / 2}}{\pi \sqrt{3}} K_{2 / 3}\left(\frac{2}{3} x^{3 / 2}\right)
\label{equation:S29}
\end{equation}

$K_i$ denotes the $i$-th order modified Bessel function of the second kind.

\begin{equation}
\operatorname{Ai}_1(\xi)=\frac{1}{\pi \sqrt{3}} \int_\rho^{\infty} d x \mathrm{~K}_{\frac{1}{3}}(x)=\frac{1}{\pi \sqrt{3}} \operatorname{IntK}_{\frac{1}{3}}(\rho),
\label{equation:S30}
\end{equation}

\begin{equation}
\frac{\operatorname{Ai}^{\prime}(\xi)}{\xi}=-\frac{1}{\pi \sqrt{3}} K_{\frac{2}{3}}(\rho),
\label{equation:S31}
\end{equation}

where $\rho=\frac{2}{3 \chi_\gamma} \frac{\omega^2}{\varepsilon_{+} \varepsilon}$.

Substituting Eqs.~\ref{equation:S25}--\ref{equation:S27} into Eq.~\ref{equation:S24} and taking an infinitesimal time step $\Delta t$ gives

\begin{widetext}
\begin{equation}
\begin{aligned}
\mathbb{P}= & \mathbb{P}^0+\mathbb{P}^{\mathrm{L}} \\
= & \frac{1}{2}\left(1-\int_0^1 d s C_p\left[\operatorname{IntK}_{\frac{1}{3}}(\rho)+\frac{s^2+(1-s)^2}{s(1-s)} \mathrm{K}_{\frac{2}{3}}(\rho)-\mathrm{K}_{\frac{2}{3}}(\rho) \hat{\mathbf{e}}_3 \cdot \boldsymbol{\xi}^i\right] \Delta t\right)+ \\
& \frac{1}{2} \boldsymbol{\xi}^f\left[\left(1-\int_0^1 d s C_p\left[\operatorname{IntK}_{\frac{1}{3}}(\rho)+\frac{s^2+(1-s)^2}{s(1-s)} \mathrm{K}_{\frac{2}{3}}(\rho)\right] \Delta t\right) \boldsymbol{\xi}^i+ \int_0^1 d s C_p \mathrm{~K}_{\frac{2}{3}}(\rho) \hat{\mathbf{e}}_3 \Delta t\right]+ \\
& \frac{\alpha m^2}{2 \omega} \Delta t \int_0^1 d s \frac{\operatorname{Gi}^{\prime}(u)}{u} \boldsymbol{\xi}^i \varepsilon_{3 i j} \boldsymbol{\xi}^f
\end{aligned}
\label{equation:S32}
\end{equation}
\end{widetext}

The expression can also be rearranged as:

\begin{equation}
\mathbb{P}=P_{V D}^L+P_{V B}^L
\label{equation:S33}
\end{equation}

where

\begin{widetext}
\begin{equation}
\begin{aligned}
P_{V D}^L= & \frac{1}{2}\left(1-\int_0^1 d s C_p\left[\operatorname{IntK}_{\frac{1}{3}}(\rho)+\frac{s^2+(1-s)^2}{s(1-s)} \mathrm{K}_{\frac{2}{3}}(\rho)-\mathrm{K}_{\frac{2}{3}}(\rho) \hat{\mathbf{e}}_3 \cdot \boldsymbol{\xi}^i\right] \Delta t\right)+ \\
& \frac{1}{2} \boldsymbol{\xi}^f\left[\left(1-\int_0^1 d s C_p\left[\operatorname{IntK}_{\frac{1}{3}}(\rho)+\frac{s^2+(1-s)^2}{s(1-s)} \mathrm{K}_{\frac{2}{3}}(\rho)\right] \Delta t\right) \boldsymbol{\xi}^i+\int_0^1 d s C_p \mathrm{~K}_{\frac{2}{3}}(\rho) \hat{\mathbf{e}}_3 \Delta t\right]
\end{aligned}
\label{equation:S34}
\end{equation}
\end{widetext}

\begin{equation}
\begin{aligned}
P_{VB}^L
 &=\frac{\alpha m^2\Delta t}{2\omega}
   \int_0^1 \mathrm{d}s\,
   \frac{\operatorname{Gi}'(u)}{u}\,
   \xi^i\varepsilon_{3ij}\xi^f \\
 &=\frac{\alpha m^2\Delta t}{2\omega}
   \int_0^1 \mathrm{d}s\,
   \frac{\operatorname{Gi}'(u)}{u}
   \left(\xi_2^i\xi_1^f-\xi_1^i\xi_2^f\right).
\end{aligned}
\label{equation:S35}
\end{equation}

Here, $C_p = \frac{\alpha m^2}{\sqrt{3} \pi \omega}$, $u = \frac{1}{\left[s(1-s) \chi_\gamma\right]^{2 / 3}}$, and $\operatorname{Gi}^{\prime}(x)$ is the derivative of the Scorer function $\operatorname{Gi}$.

\subsection{Benchmarking against literature results}


\subsubsection{Generation of polarized electrons via the collision of an elliptically polarized laser pulse with an unpolarized electron beam}

When circularly or linearly polarized laser pulses collide with unpolarized electron beams, the resulting beams exhibit zero average spin polarization due to the intrinsic symmetry of the laser field. However, for elliptically polarized (EP) laser-electron collisions, a phase difference exists between the field components governing radiation emission and the spin quantization axis relative to the transverse electron momentum. Under the influence of spin-dependent radiation reaction, the initial unpolarized electron beam splits into two distinct sub-beams characterized by opposite spin polarizations.

As a test of the implementation, we simulated the collision of an EP laser pulse with 
an unpolarized electron beam and and compared the results with those of 
Ref.~\cite{Li2019}. The laser wavelength was 
$\lambda_0=1~\mu\mathrm{m}$, the normalized amplitude was $a_0=100$, 
the pulse duration was $\tau=18T_0$, the ellipticity was $\epsilon=0.05$, 
and the temporal envelope was proportional to $\sin^2$. The initially 
unpolarized electron beam had an energy of $E_e=4$~GeV, a radius of 
$r_b=1~\mu\mathrm{m}$, and a length of $L_b=1~\mu\mathrm{m}$; initial energy 
and angular spreads were neglected. The $20\lambda_0\times15\lambda_0\times15\lambda_0$ 
domain was resolved by $960\times720\times720$ cells with a time step 
$\Delta t=0.5\Delta x/c$.

\begin{figure}[htbp]
    \centering
    \includegraphics[width=\columnwidth]{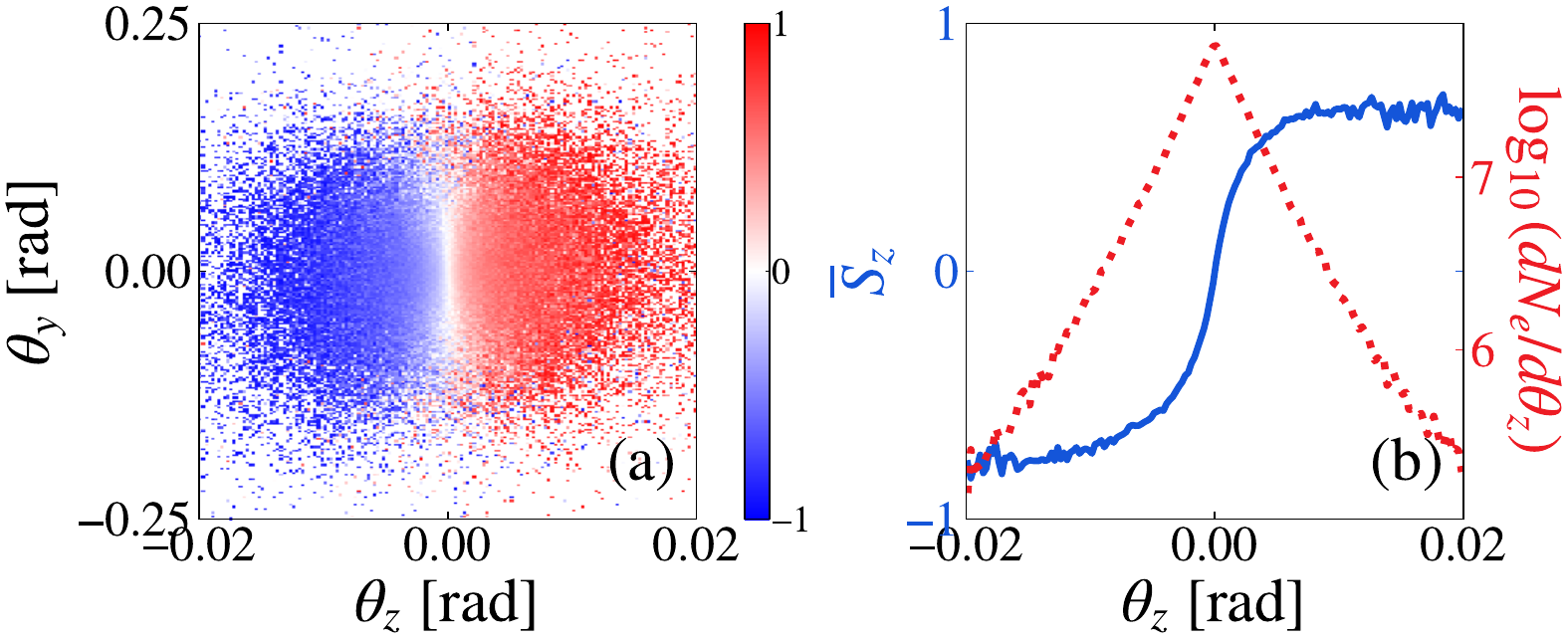}
    \caption{%
(a) Transverse distribution of the average electron spin component
$\overline{S}_z$ as a function of the deflection angles
$\theta_z=\arctan[p_z/(-p_x)]$ and
$\theta_y=\arctan[p_y/(-p_x)]$.
(b) Average spin polarization $\overline{S}_z$ (blue solid curve)
and electron angular distribution
$\log_{10}(\mathrm{d}N_e/\mathrm{d}\theta_z)$ (red dotted curve)
as functions of $\theta_z$.
}\label{figure:spinbenchmark}
\end{figure}

The electron beam splits into two sub-beams with opposite spin polarizations, in close agreement with Ref.~\cite{Li2019}; see Fig.~\ref{figure:spinbenchmark}.

\subsubsection{Dense polarized positrons from beam-solid interactions}


\begin{figure}[htbp]
    \centering
\includegraphics[width=\columnwidth]{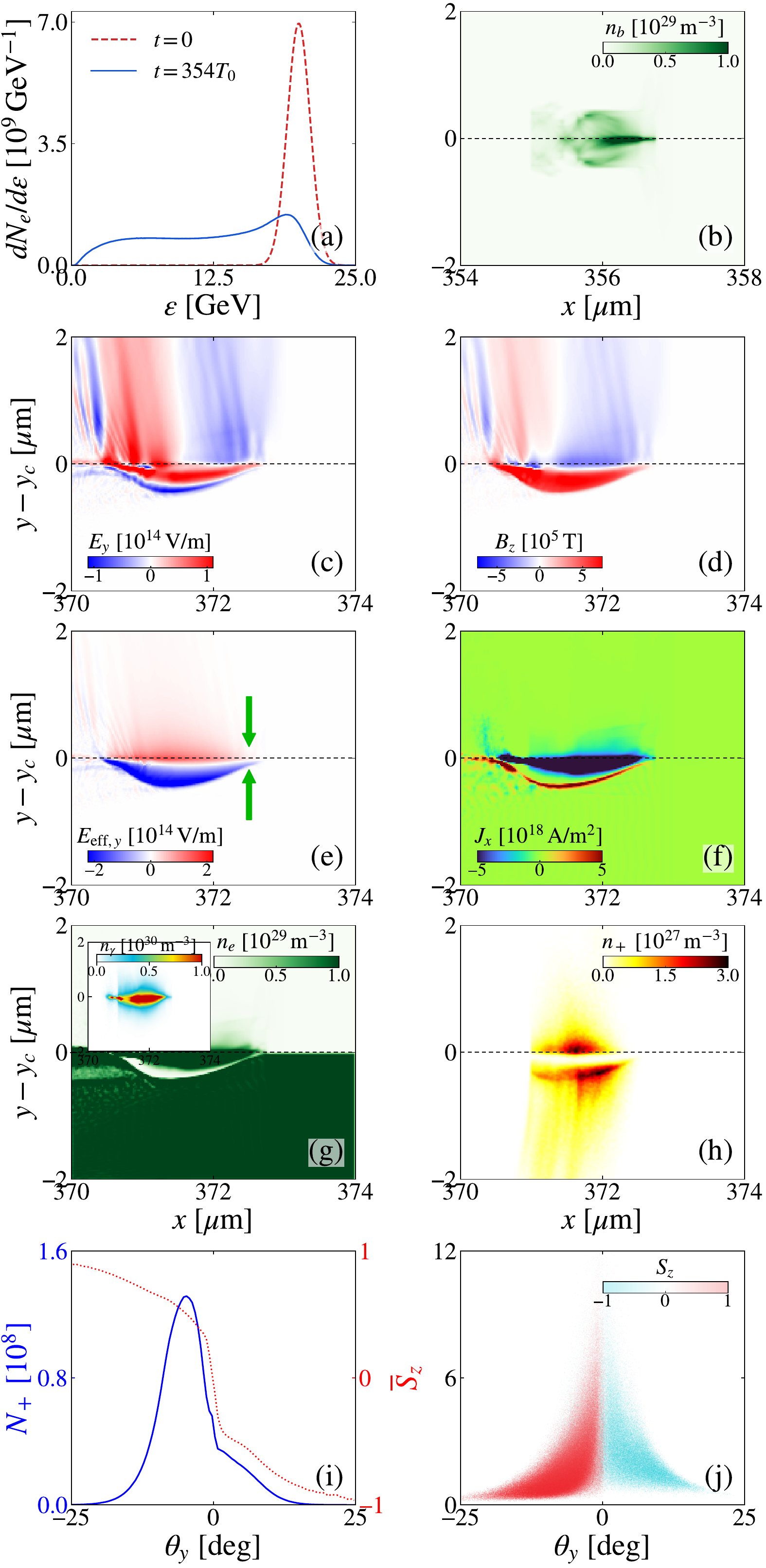}
    \caption{
Evolution of the electron beam and plasma response in the cone--converter target.
(a) Electron energy spectra at $t=0$ and $t=354T_0$.
(b) Electron-beam density $n_b$ at $t=354T_0$.
(c--h) Distributions in the $x$--$y$ plane at $z=2.5~\mu\mathrm{m}$ and $t=370T_0$: 
(c) transverse electric field $E_y$,
(d) azimuthal magnetic-field component $B_z$,
(e) effective transverse field $E_{\mathrm{eff},y}=E_y-B_z$,
(f) longitudinal current density $J_x$,
(g) total electron density $n_e$, with the photon density $n_\gamma$ shown in the inset, and
(h) positron density $n_+$.
The dashed horizontal lines indicate the converter surface at $y-y_c=0$.
(i) Angular distribution of the generated positrons (blue) and their mean longitudinal spin polarization $\overline{S}_z$ (red).
(j) Positron energy--angle distribution, with the color denoting the longitudinal spin component $S_z$.
}
\label{fig:converter_diagnostics}
\end{figure}

When a high-energy, high-density electron beam grazes the surface of a dense
plasma target, the Coulomb field of the incident beam strongly expels electrons
from the target surface. The displaced surface electrons subsequently induce a
strong return current, thereby generating a quasi-static magnetic field. In this
field, the high-energy electrons emit energetic photons via nonlinear Compton
scattering. These photons subsequently undergo the nonlinear Breit--Wheeler
(BW) process in the quasi-static magnetic field. Through spin-dependent
radiation dynamics, this interaction produces a high-density, spin-polarized
positron beam \cite{PhysRevLett.132.235001}. We reproduced this physical process
using the SPIRE code and compared our simulation results with those reported in
the literature.

3D PIC simulations were performed using the 
SPIRE code. A moving simulation window of
$4~\mu\mathrm{m}\times5~\mu\mathrm{m}\times5~\mu\mathrm{m}$
propagated along the $+x$ direction at the speed of light. 
The reference wavelength was
$\lambda_0=1~\mu\mathrm{m}$, and the domain was resolved by
$192\times240\times240$ cells, corresponding to
$\Delta x=\Delta y=\Delta z=\lambda_0/48$.
The time step was $\Delta t=T_0/96$, and the simulation lasted for
$374T_0$. Non-periodic boundary 
conditions were used in all directions. Prior to the simulation, 
the initial electromagnetic-field distributions of the high-density 
relativistic electron beam were obtained self-consistently~\cite{MASSIMO2016378}.

The electron beam propagated along $+x$ and was initially centered at
$(x_0,y_0,z_0)=(2,2.5,2.5)~\mu\mathrm{m}$. Its density profile was
\begin{equation}
\begin{aligned}
n_b(x,y,z)={}&n_{b0}\exp\!\left[
-\frac{(x-x_0)^2}{\sigma_{\parallel}^2}\right.\\
&\left.\hspace{3.5em}
-\frac{(y-y_0)^2+(z-z_0)^2}{\sigma_{\perp}^2}
\right],
\end{aligned}
\end{equation}
where $\sigma_{\parallel}=1~\mu\mathrm{m}$ and
$\sigma_{\perp}=2~\mu\mathrm{m}$. The distribution was truncated at
$1<x<3~\mu\mathrm{m}$ and within the circular transverse region
$(y-y_0)^2+(z-z_0)^2\leq(2~\mu\mathrm{m})^2$.
The particle weights were normalized after truncation 
to preserve a total beam charge of $2.8~\mathrm{nC}$. 
The beam had a mean kinetic energy of $20~\mathrm{GeV}$, 
an rms energy spread of $5\%$, and a normalized rms emittance of
$4~\mathrm{mm\,mrad}$ in both transverse directions. 
The beam was represented by 27 macroparticles per cell.

The target consisted of a hollow conical shell connected 
directly to a solid converter. The cone had a length of
$350~\mu\mathrm{m}$, entrance and exit inner diameters of
$4$ and $0.8~\mu\mathrm{m}$, respectively, and a radial wall thickness of
$0.5~\mu\mathrm{m}$. The converter had a length of $20~\mu\mathrm{m}$ 
and a semicircular transverse cross-section of radius $2~\mu\mathrm{m}$.
Both target components consisted of fully ionized carbon plasma with
$n_{e0}=89.7n_c\simeq1.0\times10^{29}~\mathrm{m^{-3}}$,
$n_{\mathrm{C}^{6+}}={n_{e0}}/{6}$.
The background electrons and $\mathrm{C}^{6+}$ ions 
were each represented by eight macroparticles per cell.


Figure~\ref{fig:converter_diagnostics} presents the electromagnetic-field
distributions, emitted-photon distributions, and spin-resolved energy spectra
of the produced positrons. These results which reproduce the main features 
reported in Figs. 2 and 3 of Ref.~\cite{PhysRevLett.132.235001}. This case 
exercises both the long-distance moving-window capability and the spin-resolved 
radiation and pair-production modules.

\section{Conclusion}

We have developed SPIRE, an integrated PIC framework for simulations of laser–plasma 
interactions with spin- and photon-polarization-resolved strong-field QED processes. 
Its classical module combines a Yee-grid FDTD field solver, a relativistic Boris 
particle pusher, particle–grid interpolation, charge-conserving current deposition, 
laser boundary injection, and a relativistic moving window in one, two, and three 
spatial dimensions. A vectorial Laguerre–Gaussian laser injector based on an 
angular-spectrum formulation further enables simulations of tightly focused, 
nonparaxial laser fields carrying optical orbital angular momentum.

The strong-field QED module treats spin- and polarization-resolved nonlinear 
Compton scattering and nonlinear Breit–Wheeler pair production within the local 
constant-field approximation. Particle spin evolves through stochastic transitions 
associated with photon emission and no-emission corrections, together with 
Thomas–Bargmann–Michel–Telegdi precession including a field-dependent 
anomalous magnetic moment. Photon polarization is described by the Stokes 
vector and evolves through nonlinear Compton scattering, nonlinear 
Breit–Wheeler pair production, vacuum birefringence, and vacuum dichroism. 
These models are incorporated into the PIC framework to describe the coupled 
evolution of plasma dynamics, radiation, particle spin, and photon polarization.

The implementation was verified through benchmark problems covering 
electrostatic expansion, laser-driven radiation, beam-driven wakefields, 
moving-window laser wakefield acceleration, and vectorial Laguerre–Gaussian 
laser injection. The classical PIC results show close agreement with 
independent SMILEI simulations and with the corresponding theoretical 
expectations. The spin- and polarization-resolved QED modules reproduce 
published results for polarized electron generation, vacuum birefringence 
and dichroism, and dense polarized positron production. Together, these 
benchmarks provide numerical verification of the individual modules and their 
coupling across different spatial dimensionalities, laser configurations, 
and strong-field QED processes. SPIRE therefore provides a numerical 
framework for studying laser–plasma interactions in which radiation 
reaction, particle spin, and photon polarization play an important role.

\bibliography{reference/references}

@article{PhysRevLett.132.235001,
  title     = {Dense Polarized Positrons from Beam-Solid Interaction},
  author    = {Zhu, Xing-Long and Liu, Wei-Yuan and Yu, Tong-Pu and Chen, Min and Weng, Su-Ming and Wang, Wei-Min and Sheng, Zheng-Ming},
  journal   = {Phys. Rev. Lett.},
  volume    = {132},
  issue     = {23},
  pages     = {235001},
  numpages  = {7},
  year      = {2024},
  month     = {Jun},
  publisher = {American Physical Society},
  doi       = {10.1103/PhysRevLett.132.235001},
  url       = {https://link.aps.org/doi/10.1103/PhysRevLett.132.235001}
}

@book{Birdsall1991Plasma,
  author    = {Birdsall, C. K. and Langdon, A. B. and Langdon, A. B.},
  title     = {Plasma Physics via Computer Simulation},
  edition   = {1},
  publisher = {CRC Press},
  year      = {1991}
}

@article{1966Numerical,
  title   = {Numerical solution of initial boundary value problems involving maxwell's equations in isotropic media},
  author  = { Yee, Kane S. },
  journal = {IEEE Transactions on Antennas \& Propagation},
  volume  = {14},
  number  = {5},
  pages   = {302-307},
  year    = {1966}
}

@book{2005Computational,
  title     = {Computational Electrodynamics: The Finite-Difference Time-Domain Method},
  author    = {Taflove, A. and Hagness, S. C.},
  edition   = {3},
  publisher = {Artech House},
  address   = {Boston},
  year      = {2005}
}

@inproceedings{Boris1970PlasmaConf,
  author    = {Boris, J. P.},
  title     = {Proceedings of 4th conference on numerical simulation of plasmas},
  publisher = {Naval Research Laboratory},
  address   = {Washington D. C.},
  year      = {1970},
  pages     = {3--67}
}

@article{ABE1986247,
  title   = {High-order spline interpolations in the particle simulation},
  journal = {Journal of Computational Physics},
  volume  = {63},
  number  = {2},
  pages   = {247-267},
  year    = {1986},
  issn    = {0021-9991},
  doi     = {https://doi.org/10.1016/0021-9991(86)90193-2},
  url     = {https://www.sciencedirect.com/science/article/pii/0021999186901932},
  author  = {Hirotade Abe and Natsuhiko Sakairi and Ryohei Itatani and Hideo Okuda}
}

@article{ESIRKEPOV2001144,
  title   = {Exact charge conservation scheme for Particle-in-Cell simulation with an arbitrary form-factor},
  journal = {Computer Physics Communications},
  volume  = {135},
  number  = {2},
  pages   = {144-153},
  year    = {2001},
  issn    = {0010-4655},
  doi     = {https://doi.org/10.1016/S0010-4655(00)00228-9},
  url     = {https://www.sciencedirect.com/science/article/pii/S0010465500002289},
  author  = {T.Zh. Esirkepov}
}

@techreport{RuhlPSCIntro,
  author      = {Ruhl, Hartmut},
  title       = {Classical Particle Simulations with the PSC code: An introduction into the PSC},
  institution = {Ruhr-Universität Bochum (RUB)},
  address     = {Bochum, Germany}
}

@article{Ritus1985,
  author  = {Ritus, V. I.},
  title   = {Quantum effects of the interaction of elementary particles with an intense electromagnetic field},
  journal = {Journal of Soviet Laser Research},
  volume  = {6},
  number  = {5},
  pages   = {497--617},
  year    = {1985},
  month   = {9},
  issn    = {1573-8760},
  doi     = {10.1007/BF01120220},
  url     = {https://doi.org/10.1007/BF01120220}
}

@article{Khokonov2010,
  author  = {Khokonov, M. Kh. and Bekulova, I. Z.},
  title   = {Length of formation of processes in a constant external field at high energies},
  journal = {Technical Physics},
  volume  = {55},
  number  = {5},
  pages   = {728--731},
  year    = {2010},
  month   = {5},
  issn    = {1090-6525},
  doi     = {10.1134/S106378421005021X},
  url     = {https://doi.org/10.1134/S106378421005021X}
}

@article{Shen2023StrongField,
  title   = {Strong-field QED effects in extreme laser-plasma interactions},
  author  = {Shen, B. F. and others},
  journal = {Physical Review Research},
  volume  = {5},
  year    = {2023}
}

@article{Torgrimsson_2021,
  journal   = {New J. Phys.},
  doi       = {10.1088/1367-2630/abf274},
  url       = {https://doi.org/10.1088/1367-2630/abf274},
  year      = 2021,
  month     = {may},
  publisher = {{IOP} Publishing},
  volume    = {23},
  number    = {6},
  pages     = {065001},
  author    = {Greger Torgrimsson},
  title     = {Loops and polarization in strong-field {QED}}
}

@article{Elkina2011,
  title     = {QED cascades induced by circularly polarized laser fields},
  author    = {Elkina, N. V. and Fedotov, A. M. and Kostyukov, I. Yu. and Legkov, M. V. and Narozhny, N. B. and Nerush, E. N. and Ruhl, H.},
  journal   = {Phys. Rev. ST Accel. Beams},
  volume    = {14},
  issue     = {5},
  pages     = {054401},
  numpages  = {12},
  year      = {2011},
  month     = {May},
  publisher = {American Physical Society},
  doi       = {10.1103/PhysRevSTAB.14.054401},
  url       = {https://link.aps.org/doi/10.1103/PhysRevSTAB.14.054401}
}

@misc{CAIN1,
  note   = {\url{ http://lcdev.kek.jp/~yokoya/CAIN/}},
  author = {User’s Manual of CAIN Version 2.42}
}

@article{Bargmann1959,
  title     = {Precession of the Polarization of Particles Moving in a Homogeneous Electromagnetic Field},
  author    = {Bargmann, V. and Michel, Louis and Telegdi, V. L.},
  journal   = {Phys. Rev. Lett.},
  volume    = {2},
  issue     = {10},
  pages     = {435--436},
  numpages  = {0},
  year      = {1959},
  month     = {May},
  publisher = {American Physical Society},
  doi       = {10.1103/PhysRevLett.2.435},
  url       = {https://link.aps.org/doi/10.1103/PhysRevLett.2.435}
}

@article{Gonoskov2015,
  title     = {Extended particle-in-cell schemes for physics in ultrastrong laser fields: Review and developments},
  author    = {Gonoskov, A. and Bastrakov, S. and Efimenko, E. and Ilderton, A. and Marklund, M. and Meyerov, I. and Muraviev, A. and Sergeev, A. and Surmin, I. and Wallin, E.},
  journal   = {Phys. Rev. E},
  volume    = {92},
  issue     = {2},
  pages     = {023305},
  numpages  = {18},
  year      = {2015},
  month     = {Aug},
  publisher = {American Physical Society},
  doi       = {10.1103/PhysRevE.92.023305},
  url       = {https://link.aps.org/doi/10.1103/PhysRevE.92.023305}
}

@book{Berestetskii1982,
  title     = {Quantum Electrodynamics},
  author    = {V. B. Berestetskii and E. M. Lifshitz and L. P. Pitaevskii},
  year      = 1982,
  publisher = {Pergamon},
  address   = {Oxford}
}

@article{Meuren2011,
  title     = {Quantum Electron Self-Interaction in a Strong Laser Field},
  author    = {Meuren, S. and Di Piazza, A.},
  journal   = {Phys. Rev. Lett.},
  volume    = {107},
  issue     = {26},
  pages     = {260401},
  numpages  = {5},
  year      = {2011},
  month     = {Dec},
  publisher = {American Physical Society},
  doi       = {10.1103/PhysRevLett.107.260401},
  url       = {https://link.aps.org/doi/10.1103/PhysRevLett.107.260401}
}

@article{Baier1972,
  author  = {Ba\ifmmode \check{i}\else \v{i}\fi{}er, V. N.},
  title   = {RADIATIVE POLARIZATION OF ELECTRONS IN STORAGE RINGS},
  journal = {Sov. Phys. Usp.},
  volume  = {14},
  number  = {6},
  pages   = {695},
  url     = {http://stacks.iop.org/0038-5670/14/i=6/a=R02},
  year    = {1972}
}

@misc{CAIN,
  note = {User’s Manual of CAIN: \url{https://www-jlc.kek.jp/subg/ir/lib/cain21b.manual/},
          }
}

@article{Ilderton2020,
  title     = {Loop spin effects in intense background fields},
  author    = {Ilderton, A. and King, B. and Tang, S.},
  journal   = {Phys. Rev. D},
  volume    = {102},
  issue     = {7},
  pages     = {076013},
  numpages  = {10},
  year      = {2020},
  month     = {Oct},
  publisher = {American Physical Society},
  doi       = {10.1103/PhysRevD.102.076013},
  url       = {https://link.aps.org/doi/10.1103/PhysRevD.102.076013}
}

@article{Bragin2017,
  title     = {High-Energy Vacuum Birefringence and Dichroism in an Ultrastrong Laser Field},
  author    = {Bragin, Sergey and Meuren, Sebastian and Keitel, Christoph H. and Di Piazza, Antonino},
  journal   = {Phys. Rev. Lett.},
  volume    = {119},
  issue     = {25},
  pages     = {250403},
  numpages  = {7},
  year      = {2017},
  month     = {Dec},
  publisher = {American Physical Society},
  doi       = {10.1103/PhysRevLett.119.250403},
  url       = {https://link.aps.org/doi/10.1103/PhysRevLett.119.250403}
}

@article{Gonoskov2022,
  title     = {Charged particle motion and radiation in strong electromagnetic fields},
  author    = {Gonoskov, A. and Blackburn, T. G. and Marklund, M. and Bulanov, S. S.},
  journal   = {Rev. Mod. Phys.},
  volume    = {94},
  issue     = {4},
  pages     = {045001},
  numpages  = {63},
  year      = {2022},
  month     = {Oct},
  publisher = {American Physical Society},
  doi       = {10.1103/RevModPhys.94.045001},
  url       = {https://link.aps.org/doi/10.1103/RevModPhys.94.045001}
}

@article{Dai2024,
  title     = {Fermionic signal of vacuum polarization in strong laser fields},
  author    = {Dai, Ya-Nan and Hatsagortsyan, Karen Z. and Keitel, Christoph H. and Chen, Yue-Yue},
  journal   = {Phys. Rev. D},
  volume    = {110},
  issue     = {1},
  pages     = {012008},
  numpages  = {19},
  year      = {2024},
  month     = {Jul},
  publisher = {American Physical Society},
  doi       = {10.1103/PhysRevD.110.012008},
  url       = {https://link.aps.org/doi/10.1103/PhysRevD.110.012008}
}

@article{Zhuang2023,
  title     = {Laser-driven lepton polarization in the quantum radiation-dominated reflection regime},
  author    = {Zhuang, Kai-Hong and Chen, Yue-Yue and Li, Yan-Fei and Hatsagortsyan, Karen Z. and Keitel, Christoph H.},
  journal   = {Phys. Rev. D},
  volume    = {108},
  issue     = {3},
  pages     = {033001},
  numpages  = {17},
  year      = {2023},
  month     = {Aug},
  publisher = {American Physical Society},
  doi       = {10.1103/PhysRevD.108.033001},
  url       = {https://link.aps.org/doi/10.1103/PhysRevD.108.033001}
}

@article{FEDOTOV2023,
  title    = {Advances in QED with intense background fields},
  journal  = {Physics Reports},
  volume   = {1010},
  pages    = {1-138},
  year     = {2023},
  note     = {Advances in QED with intense background fields},
  issn     = {0370-1573},
  doi      = {https://doi.org/10.1016/j.physrep.2023.01.003},
  url      = {https://www.sciencedirect.com/science/article/pii/S0370157323000352},
  author   = {A. Fedotov and A. Ilderton and F. Karbstein and B. King and D. Seipt and H. Taya and G. Torgrimsson}
}

@article{McMaster1961,
  title     = {Matrix Representation of Polarization},
  author    = {McMaster, William H.},
  journal   = {Rev. Mod. Phys.},
  volume    = {33},
  issue     = {1},
  pages     = {8--28},
  numpages  = {0},
  year      = {1961},
  month     = {Jan},
  publisher = {American Physical Society},
  doi       = {10.1103/RevModPhys.33.8},
  url       = {https://link.aps.org/doi/10.1103/RevModPhys.33.8}
}

@article{Liyf2020,
  title     = {Polarized Ultrashort Brilliant Multi-GeV $\ensuremath{\gamma}$ Rays via Single-Shot Laser-Electron Interaction},
  author    = {Li, Yan-Fei and Shaisultanov, Rashid and Chen, Yue-Yue and Wan, Feng and Hatsagortsyan, Karen Z. and Keitel, Christoph H. and Li, Jian-Xing},
  journal   = {Phys. Rev. Lett.},
  volume    = {124},
  issue     = {1},
  pages     = {014801},
  numpages  = {7},
  year      = {2020},
  month     = {Jan},
  publisher = {American Physical Society},
  doi       = {10.1103/PhysRevLett.124.014801},
  url       = {https://link.aps.org/doi/10.1103/PhysRevLett.124.014801}
}

@article{Piazza2012,
  title     = {Extremely high-intensity laser interactions with fundamental quantum systems},
  author    = {Di Piazza, A. and M\"uller, C. and Hatsagortsyan, K. Z. and Keitel, C. H.},
  journal   = {Rev. Mod. Phys.},
  volume    = {84},
  issue     = {3},
  pages     = {1177--1228},
  numpages  = {0},
  year      = {2012},
  month     = {Aug},
  publisher = {American Physical Society},
  doi       = {10.1103/RevModPhys.84.1177},
  url       = {https://link.aps.org/doi/10.1103/RevModPhys.84.1177}
}

@book{Baier1998,
  author    = {V. N. Baier and V. M. Katkov and V. M. Strakhovenko},
  year      = {1998},
  title     = {Electromagnetic Processes at High Energies in Oriented Single Crystals},
  publisher = {World Scientific},
  address   = {Singapore}
}

@article{Wang2025,
  author  = {Wang, Wei-Min and Gao, Zhe and Song, Huai-Hang and Sheng, Zheng-Ming and Li, Yu-Tong and Zhang, Jie},
  title   = {QED-PIC code development and applications in QED-dominant laser plasma interactions},
  journal = {Reviews of Modern Plasma Physics},
  year    = {2025},
  month   = {sep},
  volume  = {9},
  number  = {1},
  pages   = {28},
  issn    = {2367-3192},
  doi     = {10.1007/s41614-025-00203-9},
  url     = {https://doi.org/10.1007/s41614-025-00203-9}
}

@article{Courant1928,
  author  = {Courant, R. and Friedrichs, K. and Lewy, H.},
  title   = {{\"U}ber die partiellen Differenzengleichungen der mathematischen Physik},
  journal = {Mathematische Annalen},
  year    = {1928},
  month   = {dec},
  volume  = {100},
  number  = {1},
  pages   = {32--74},
  issn    = {1432-1807},
  url     = {https://doi.org/10.1007/BF01448839}
}

@article{Li2019,
  title     = {Ultrarelativistic Electron-Beam Polarization in Single-Shot Interaction with an Ultraintense Laser Pulse},
  author    = {Li, Yan-Fei and Shaisultanov, Rashid and Hatsagortsyan, Karen Z. and Wan, Feng and Keitel, Christoph H. and Li, Jian-Xing},
  journal   = {Phys. Rev. Lett.},
  volume    = {122},
  issue     = {15},
  pages     = {154801},
  numpages  = {6},
  year      = {2019},
  month     = {Apr},
  publisher = {American Physical Society},
  doi       = {10.1103/PhysRevLett.122.154801},
  url       = {https://link.aps.org/doi/10.1103/PhysRevLett.122.154801}
}

@article{Allen1992OAM,
  author  = {Allen, L. and Beijersbergen, M. W. and
             Spreeuw, R. J. C. and Woerdman, J. P.},
  title   = {Orbital Angular Momentum of Light and the Transformation
             of Laguerre--Gaussian Laser Modes},
  journal = {Physical Review A},
  volume  = {45},
  number  = {11},
  pages   = {8185--8189},
  year    = {1992},
  doi     = {10.1103/PhysRevA.45.8185}
}

@article{RichardsWolf1959,
  author  = {Richards, B. and Wolf, E.},
  title   = {Electromagnetic Diffraction in Optical Systems.
             {II}. Structure of the Image Field in an Aplanatic System},
  journal = {Proceedings of the Royal Society of London.
             Series A, Mathematical and Physical Sciences},
  volume  = {253},
  number  = {1274},
  pages   = {358--379},
  year    = {1959},
  doi     = {10.1098/rspa.1959.0200}
}

@article{Yee1966,
  author  = {Yee, K. S.},
  title   = {Numerical Solution of Initial Boundary Value Problems
             Involving {Maxwell's} Equations in Isotropic Media},
  journal = {IEEE Transactions on Antennas and Propagation},
  volume  = {14},
  number  = {3},
  pages   = {302--307},
  year    = {1966},
  month   = may,
  doi     = {10.1109/TAP.1966.1138693}
}

@article{UmashankarTaflove1982,
  author  = {Umashankar, K. R. and Taflove, A.},
  title   = {A Novel Method to Analyze Electromagnetic Scattering
             of Complex Objects},
  journal = {IEEE Transactions on Electromagnetic Compatibility},
  volume  = {EMC-24},
  number  = {4},
  pages   = {397--405},
  year    = {1982},
  month   = nov,
  doi     = {10.1109/TEMC.1982.304054}
}

@article{Wan2023,
  author  = {Wan, Feng and Lv, Chong and Xue, Kun and Dou, Zhen-Ke and Zhao, Qian and Ababekri, Mamutjan and Wei, Wen-Qing and Li, Zhong-Peng and Zhao, Yong-Tao and Li, Jian-Xing},
  title   = {Simulations of spin/polarization-resolved laser–plasma interactions in the nonlinear QED regime},
  journal = {Matter and Radiation at Extremes},
  volume  = {8},
  number  = {6},
  pages   = {064002},
  year    = {2023},
  month   = {09},
  issn    = {2468-2047},
  doi     = {10.1063/5.0163929},
  url     = {https://doi.org/10.1063/5.0163929}
}

@article{Wang2015,
  title     = {Integrated simulation approach for laser-driven fast ignition},
  author    = {Wang, W.-M. and Gibbon, P. and Sheng, Z.-M. and Li, Y.-T.},
  journal   = {Phys. Rev. E},
  volume    = {91},
  issue     = {1},
  pages     = {013101},
  numpages  = {10},
  year      = {2015},
  month     = {Jan},
  publisher = {American Physical Society},
  doi       = {10.1103/PhysRevE.91.013101},
  url       = {https://link.aps.org/doi/10.1103/PhysRevE.91.013101}
}

@article{DEROUILLAT2018351,
  title   = {Smilei : A collaborative, open-source, multi-purpose particle-in-cell code for plasma simulation},
  journal = {Computer Physics Communications},
  volume  = {222},
  pages   = {351-373},
  year    = {2018},
  issn    = {0010-4655},
  doi     = {https://doi.org/10.1016/j.cpc.2017.09.024},
  url     = {https://www.sciencedirect.com/science/article/pii/S0010465517303314},
  author  = {J. Derouillat and A. Beck and F. Pérez and T. Vinci and M. Chiaramello and A. Grassi and M. Flé and G. Bouchard and I. Plotnikov and N. Aunai and J. Dargent and C. Riconda and M. Grech}
}

@misc{YUNIC,
  title         = {YUNIC: A Multi-Dimensional Particle-In-Cell Code for Laser-Plasma Interaction},
  author        = {Huai-Hang Song and Wei-Min Wang and Yu-Tong Li},
  year          = {2021},
  eprint        = {2104.00642},
  archiveprefix = {arXiv},
  primaryclass  = {physics.plasm-ph},
  url           = {https://arxiv.org/abs/2104.00642}
}

@article{PhysRevE.77.046405,
  title     = {Single-cycle powerful megawatt to gigawatt terahertz pulse radiated from a wavelength-scale plasma oscillator},
  author    = {Wu, Hui-Chun and Sheng, Zheng-Ming and Zhang, Jie},
  journal   = {Phys. Rev. E},
  volume    = {77},
  issue     = {4},
  pages     = {046405},
  numpages  = {6},
  year      = {2008},
  month     = {Apr},
  publisher = {American Physical Society},
  doi       = {10.1103/PhysRevE.77.046405},
  url       = {https://link.aps.org/doi/10.1103/PhysRevE.77.046405}
}

@article{Song2021BreitWheeler,
  author  = {Song, Huai-Hang and Wang, Wei-Min and Li, Yan-Fei
             and Li, Bing-Jun and Li, Yu-Tong and Sheng, Zheng-Ming
             and Chen, Li-Ming and Zhang, Jie},
  title   = {Spin and polarization effects on the nonlinear
             {Breit--Wheeler} pair production in laser-plasma interaction},
  journal = {New Journal of Physics},
  volume  = {23},
  number  = {7},
  pages   = {075005},
  year    = {2021},
  doi     = {10.1088/1367-2630/ac0dec},
  url     = {https://doi.org/10.1088/1367-2630/ac0dec}
}

@article{MASSIMO2016378,
  title   = {Electromagnetic self-consistent field initialization and fluid advance techniques for hybrid-kinetic PWFA code Architect},
  journal = {Nuclear Instruments and Methods in Physics Research Section A: Accelerators, Spectrometers, Detectors and Associated Equipment},
  volume  = {829},
  pages   = {378-382},
  year    = {2016},
  note    = {2nd European Advanced Accelerator Concepts Workshop - EAAC 2015},
  issn    = {0168-9002},
  doi     = {https://doi.org/10.1016/j.nima.2016.02.043},
  url     = {https://www.sciencedirect.com/science/article/pii/S0168900216002102},
  author  = {F. Massimo and A. Marocchino and A.R. Rossi}
}

@article{Thiele2016TightFocus,
  author  = {Thiele, Illia and Skupin, Stefan and Nuter, Rachel},
  title   = {Boundary conditions for arbitrarily shaped and tightly
             focused laser pulses in electromagnetic codes},
  journal = {Journal of Computational Physics},
  volume  = {321},
  pages   = {1110--1119},
  year    = {2016},
  doi     = {10.1016/j.jcp.2016.06.004}
}

@article{Capoglu2008FocusedFDTD,
  author  = {{\c{C}}apo{\u{g}}lu, {\.I}lker R. and
             Taflove, Allen and Backman, Vadim},
  title   = {Generation of an incident focused light pulse in {FDTD}},
  journal = {Optics Express},
  volume  = {16},
  number  = {23},
  pages   = {19208--19220},
  year    = {2008},
  doi     = {10.1364/OE.16.019208}
}

@article{Charbonnet2025FocusedPIC,
  author  = {Charbonnet, K. J. and Nelson, E. C. and
             Reutershan, T. and Barty, C. P. J.},
  title   = {Computational methods for focused arbitrary laser fields
             in plasma simulations},
  journal = {Physics of Plasmas},
  volume  = {32},
  number  = {4},
  pages   = {043901},
  year    = {2025},
  doi     = {10.1063/5.0250718}
}

@article{Gonzalez2018MaxwellConsistent,
  author  = {Gonz{\'a}lez de Alaiza Mart{\'i}nez, P. and
             Duchateau, G. and Chimier, B. and Nuter, R. and
             Thiele, I. and Skupin, S. and Tikhonchuk, V. T.},
  title   = {Maxwell-consistent, symmetry- and energy-preserving
             solutions for ultrashort-laser-pulse propagation beyond
             the paraxial approximation},
  journal = {Physical Review A},
  volume  = {98},
  number  = {4},
  pages   = {043849},
  year    = {2018},
  doi     = {10.1103/PhysRevA.98.043849}
}

@article{KimLee1999NonparaxialLG,
  author  = {Kim, H. C. and Lee, Y. H.},
  title   = {Hermite--Gaussian and Laguerre--Gaussian beams beyond
             the paraxial approximation},
  journal = {Optics Communications},
  volume  = {169},
  number  = {1--6},
  pages   = {9--16},
  year    = {1999},
  doi     = {10.1016/S0030-4018(99)00411-3}
}

@article{Chen1985PWFA,
  author  = {Chen, Pisin and Dawson, J. M. and Huff, Robert W. and
             Katsouleas, T.},
  title   = {Acceleration of Electrons by the Interaction of a Bunched
             Electron Beam with a Plasma},
  journal = {Physical Review Letters},
  volume  = {54},
  number  = {7},
  pages   = {693--696},
  year    = {1985},
  doi     = {10.1103/PhysRevLett.54.693}
}

@article{Lu2006Blowout,
  author  = {Lu, W. and Huang, C. and Zhou, M. and Mori, W. B. and
             Katsouleas, T.},
  title   = {Nonlinear Theory for Relativistic Plasma Wakefields in the
             Blowout Regime},
  journal = {Physical Review Letters},
  volume  = {96},
  number  = {16},
  pages   = {165002},
  year    = {2006},
  doi     = {10.1103/PhysRevLett.96.165002}
}

@article{Blumenfeld2007EnergyDoubling,
  author  = {Blumenfeld, I. and Clayton, C. E. and Decker, F.-J. and others},
  title   = {Energy Doubling of 42 {GeV} Electrons in a Metre-Scale
             Plasma Wakefield Accelerator},
  journal = {Nature},
  volume  = {445},
  pages   = {741--744},
  year    = {2007},
  doi     = {10.1038/nature05538}
}

@article{Caldwell2009ProtonPWFA,
  author  = {Caldwell, Allen and Lotov, Konstantin and Pukhov, Alexander
             and Simon, Frank},
  title   = {Proton-Driven Plasma-Wakefield Acceleration},
  journal = {Nature Physics},
  volume  = {5},
  pages   = {363--367},
  year    = {2009},
  doi     = {10.1038/nphys1248}
}

@article{Adli2018AWAKE,
  author        = {Adli, E. and others},
  collaboration = {AWAKE Collaboration},
  title         = {Acceleration of Electrons in the Plasma Wakefield of a
                   Proton Bunch},
  journal       = {Nature},
  volume        = {561},
  pages         = {363--367},
  year          = {2018},
  doi           = {10.1038/s41586-018-0485-4}
}

@article{TajimaDawson1979,
  author  = {Tajima, T. and Dawson, J. M.},
  title   = {Laser Electron Accelerator},
  journal = {Physical Review Letters},
  volume  = {43},
  number  = {4},
  pages   = {267--270},
  year    = {1979},
  doi     = {10.1103/PhysRevLett.43.267}
}

@article{Pukhov2002LWFA,
  author  = {Pukhov, A. and Meyer-ter-Vehn, J.},
  title   = {Laser Wake Field Acceleration: The Highly Non-Linear
             Broken-Wave Regime},
  journal = {Applied Physics B},
  volume  = {74},
  number  = {4--5},
  pages   = {355--361},
  year    = {2002},
  doi     = {10.1007/s003400200795}
}

@article{Mangles2004LWFA,
  author  = {Mangles, S. P. D. and Murphy, C. D. and Najmudin, Z. and others},
  title   = {Monoenergetic Beams of Relativistic Electrons from Intense
             Laser--Plasma Interactions},
  journal = {Nature},
  volume  = {431},
  pages   = {535--538},
  year    = {2004},
  doi     = {10.1038/nature02939}
}

@article{Geddes2004LWFA,
  author  = {Geddes, C. G. R. and Toth, Cs. and van Tilborg, J. and others},
  title   = {High-Quality Electron Beams from a Laser Wakefield
             Accelerator Using Plasma-Channel Guiding},
  journal = {Nature},
  volume  = {431},
  pages   = {538--541},
  year    = {2004},
  doi     = {10.1038/nature02900}
}

@article{Faure2004LWFA,
  author  = {Faure, J. and Glinec, Y. and Pukhov, A. and others},
  title   = {A Laser--Plasma Accelerator Producing Monoenergetic
             Electron Beams},
  journal = {Nature},
  volume  = {431},
  pages   = {541--544},
  year    = {2004},
  doi     = {10.1038/nature02963}
}

@article{RIDGERS2014273,
  title   = {Modelling gamma-ray photon emission and pair production in high-intensity laser–matter interactions},
  journal = {Journal of Computational Physics},
  volume  = {260},
  pages   = {273-285},
  year    = {2014},
  issn    = {0021-9991},
  doi     = {https://doi.org/10.1016/j.jcp.2013.12.007},
  url     = {https://www.sciencedirect.com/science/article/pii/S0021999113008061},
  author  = {C.P. Ridgers and J.G. Kirk and R. Duclous and T.G. Blackburn and C.S. Brady and K. Bennett and T.D. Arber and A.R. Bell}
}

@article{PhysRevE.92.023305,
  title     = {Extended particle-in-cell schemes for physics in ultrastrong laser fields: Review and developments},
  author    = {Gonoskov, A. and Bastrakov, S. and Efimenko, E. and Ilderton, A. and Marklund, M. and Meyerov, I. and Muraviev, A. and Sergeev, A. and Surmin, I. and Wallin, E.},
  journal   = {Phys. Rev. E},
  volume    = {92},
  issue     = {2},
  pages     = {023305},
  numpages  = {18},
  year      = {2015},
  month     = {Aug},
  publisher = {American Physical Society},
  doi       = {10.1103/PhysRevE.92.023305},
  url       = {https://link.aps.org/doi/10.1103/PhysRevE.92.023305}
}

\end{document}